\documentclass{aa}
\usepackage{graphicx}
\usepackage{txfonts}
\usepackage{natbib}
\bibpunct{(}{)}{;}{a}{}{,}
\usepackage{longtable}
\usepackage{lscape}
\usepackage{mathrsfs}

\begin{document}
\title{A Sino-German $\lambda$6\ cm polarization survey of the Galactic plane}
\subtitle{II. The region from 129$\degr$ to 230$\degr$ longitude}
\author{X. Y. Gao\inst{1,2},  W. Reich\inst{2}, J. L. Han\inst{1}, X. H. Sun\inst{1}, R. Wielebinski\inst{2}, \\ W. B. Shi\inst{1,3}, L. Xiao\inst{1}, 
P. Reich\inst{2}, E. F\"urst\inst{2}, M. Z. Chen\inst{4}, J. Ma\inst{4}}

\offprints{W. Reich\\ \email{wreich@mpifr-bonn.mpg.de}}

\institute{National Astronomical Observatories, CAS, Jia-20 Datun Road, Chaoyang District, Beijing 100012, China
      \and Max-Planck-Institut f\"{u}r Radioastronomie, Auf dem H\"{u}gel 69, 
            53121 Bonn, Germany
     \and School of Space Science and Physics, Shandong University at Weihai, 180 Cultural West Road, Shandong 264209, China
  \and Urumqi Observatory, National Astronomical Observatories, CAS, 40-5 South Beijing Road, Urumqi 830011, China}

\date{Received ; accepted }

\abstract 
{Linearly polarized Galactic synchrotron emission provides valuable information about the properties of the Galactic magnetic field 
and the interstellar magneto-ionic medium, when Faraday rotation along the line of sight is properly taken into account.
}
{We aim to survey the Galactic plane at $\lambda$6\ cm including linear polarization.  
At such a short wavelength Faraday rotation effects are in general small and the Galactic magnetic field properties can be probed to larger distances than at long wavelengths.
}
{The Urumqi 25-m telescope is used for a sensitive $\lambda$6\ cm survey in total and polarized intensities. 
WMAP K-band (22.8~GHz) polarization data are used to restore the absolute zero-level of the Urumqi $U$ and  $Q$ maps by extrapolation.
}
{Total intensity and polarization maps are presented for a Galactic plane region of $129\degr \leq \ell \leq 230\degr$ and $|b| \leq 5\degr$
in the anti-centre with an angular resolution of $9\farcm5$ and an average sensitivity of 0.6~mK and 0.4~mK $\rm T_{B}$ in total and polarized intensity, 
respectively. We briefly discuss the properties of some extended Faraday Screens   
detected in the $\lambda$6\ cm polarization maps.
}
{The Sino-German $\lambda$6\ cm polarization survey provides new information about the properties of the magnetic ISM.
The survey also adds valuable information for discrete Galactic objects 
and is in particular suited to detect extended Faraday Screens with large rotation measures
hosting strong regular magnetic fields.}


\titlerunning{A Sino-German $\lambda$6\ cm polarization survey of the Galactic plane II.}
\authorrunning{X. Y. Gao et al.}

\maketitle
\section{Introduction}

Surveys of the Galactic plane at several frequencies are required to disentangle 
the individual star formation complexes, or thermal \ion{H}{II} regions, non-thermal supernova remnants (SNRs) and 
extragalactic sources.  The diffuse emission associated with the Galactic disk
is produced by relativistic electrons spiraling in magnetic fields and by thin ionized thermal gas. Both the diffuse non-thermal
emission and the SNRs have significant linear polarization. Mapping of the Galactic plane at several radio frequencies including
linear polarization offers a method to separate these non-thermal components as well as allowing a delineation 
of the Galactic magnetic field. 

The Galactic plane has been surveyed from 22~MHz up to 10~GHz, albeit usually without polarization measurements. 
Sensitive Galactic polarization plane surveys began in the 1980s. A 2.7~GHz survey using the Effelsberg 100-m telescope by \citet{Junkes87} showed a
section of the Galactic plane with $4\farcm3$ angular resolution. Further Northern sky Galactic plane surveys at 2.7~GHz 
\citep{Reich9011,Fuerst90,Duncan99} were complemented by 2.4~GHz Southern Galactic plane surveys using the Parkes 64-m telescope 
\citep{Duncan95,Duncan97}. To achieve angular resolution of arc minutes at  lower frequencies synthesis radio telescopes had
to be used for surveys: e.g. the Westerbork Synthesis Radio Telescope at 350~MHz \citep{Wieringa93,Haverkorn031,Haverkorn032}, 
the Dominion Radio Astrophysical Observatory synthesis telescope at 408~MHz and 1.4~GHz (Canadian Galactic Plane 
Survey, CGPS) \citep{Taylor03}, and the Australian Telescope Compact Array at 1.4~GHz (Southern Galactic Plane Survey, SGPS) \citep{Gaensler01,Haverkorn06}. Most of the mentioned surveys only cover a
narrow strip along the Galactic plane. To overcome this deficiency the Galactic plane was mapped at 1.4~GHz with
the Effelsberg 100-m telescope for $|b| \leq 20\degr$. First maps from this survey were shown by \citet{Uyaniker99} and by 
\citet{Reich04}. To study the nature of sources and the properties of the magnetic field, polarization surveys at higher radio frequencies are needed.
Valuable information about diffuse polarized Galactic emission was provided by WMAP 
at 22.8~GHz and higher frequencies \citep{Hinshaw09}, although the angular resolution of $50\arcmin$ at 
22.8~GHz is in general too coarse to resolve the complex Galactic structures in the Galactic plane.  

The Sino-German $\lambda$6\ cm survey, covering a 10$\degr$ wide strip of the Galactic plane, has been carried out 
since 2004 using the 25-m radio telescope of the Urumqi Observatory, National Astronomical Observatories, CAS. 
This survey fills the existing gap in frequency 
coverage by providing maps of the Galactic plane from $10\degr \leq \ell \leq 230\degr$ 
 and $|b| \leq 5\degr$ with an angular resolution of $9\farcm5$.  
 The survey maps and a list of compact sources will be released after completion of the
 $\lambda$6\ cm survey project expected for the end of 2010.
The first results have already been presented by \citet{Sun07} (hereafter called Paper~I), including details of the survey concept, the observing and 
calibration methods and the reduction process. In Paper~I, covering the longitude range from $122\degr$ to $129\degr$,
we illustrated the scientific potential provided by the $\lambda$6\ cm survey by delineating new faint 
\ion{H}{II} regions, studied spectra of SNRs, discovered Faraday Screens as well as traced the magnetic fields in this section of the Galactic plane. 
Most remarkable discoveries are two extended Faraday Screens located at the Perseus arm. One of them is caused by 
a previously unknown faint \ion{H}{II} region. Both Faraday Screens host strong regular magnetic fields with rotation measures ($RM$) of the order
of 200~rad\ m$^{-2}$. They are not visible at low frequencies because such high $RM$s cause a
polarization angle rotation by more than $180\degr$, or they are beyond the polarization horizon.
This proves the value of a sensitive $\lambda$6\ cm polarization survey to detect them in the magnetized interstellar medium. 
The commonly adopted picture of the Galactic magnetic field in the thin disk to consist of a regular component following
basically the spiral arms of the Galaxy together with a turbulent magnetic field component of about similar strength might be modified 
in case numerous extended Faraday Screens with a uniform regular magnetic field exist. The origin of such magnetic bubbles acting as Faraday Screens is not clear so far.   
 
Here we present the second section of the $\lambda$6\ cm survey for the outer Galaxy covering the region 
$129\degr \leq \ell \leq 230\degr$. 
In Sect.~2 observation and data processing details for this survey area are discussed. In Sect.~3 we present 
the total power and polarization maps (Sect.~3.1), followed by a brief discussion on the survey's potential to study
and detect SNRs (Sect.~3.2) and  \ion{H}{II} regions (Sect.~3.3), while in Sect.~3.4 we focus  on
newly detected and prominent Faraday Screens in the interstellar medium. Results are summarized in Sect.~4.

\section{Observations and Data reduction}

\subsection{Observation set-up}

The Sino-German $\lambda$6\ cm polarization survey of the Galactic plane was carried out with the 25-m Urumqi telescope.
The $\lambda$6\ cm system is a copy of an Effelsberg single-channel receiver and has a system temperature of about 22~K when pointing to the 
zenith at clear sky. The half-power beam
width (HPBW) of the telescope was $9\farcm5$. Survey observations were exclusively made during clear sky at night time. 
In ``broad band mode'' the centre frequency was 4800~MHz with a bandwidth 600~MHz, while in ``narrow band mode'' the centre frequency 
was  4963~MHz with a bandwidth of 295~MHz. 
``Narrow band mode'' was used to avoid contamination by the geostationary Indian INSAT-satellites located at four positions in southern and 
western directions, which 
emit strong signals below frequencies of about 4810~MHz. Thus all observations close to the satellite positions were made in ``narrow band mode'',
while the ``broad band mode'' was used for all other directions.
The survey region was limited at $\ell = 230\degr$, because regions with larger $\ell$
have to be observed at very low elevations, so that the increased ground radiation can not be subtracted with sufficient accuracy.
Measurements of  the ground radiation properties of the Urumqi 25-m telescope at $\lambda$6\ cm were already reported by \citet{WC07}.
The main observational survey parameters are listed in Table~1.

Survey maps were observed in two orthogonal directions: along $\ell$ and $b$. The combined survey consists of a large number of individual 
8$\degr \times 2\fdg6$ or $8\fdg2 \times 2\fdg6$ maps observed in $\ell$ direction and of  $2\degr \times 10\degr$ or $2\fdg2 \times 
10\degr$ maps observed along $b$.
Total intensity, Stokes $I$, and the linearly polarized components, Stokes $U$ and $Q$, were measured simultaneously.
3C286 and 3C295 served as the main polarized and un-polarized calibration sources, while 3C48, 3C138 served as secondary polarized 
calibrators. Calibration sources were always observed before starting a survey map in the same observation mode.

\begin{figure}
\begin{center}
\includegraphics[angle=-90,width=0.46\textwidth]{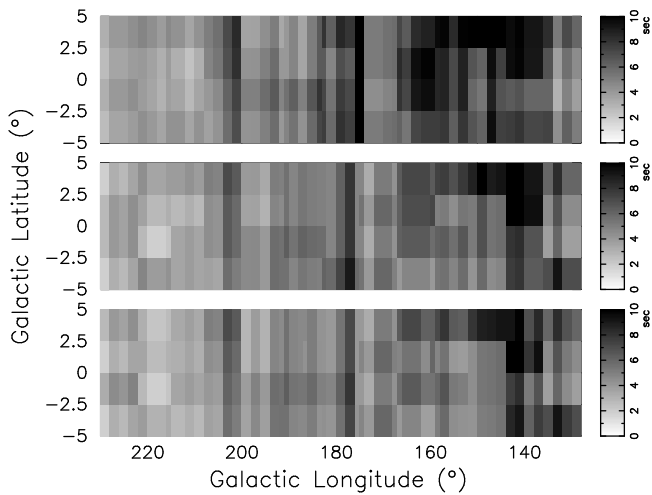}
\caption{Distribution of effective integration time in sec for the survey sub-areas shown for Stokes $I$, $U$ and $Q$ from top to bottom.}
\label{integration_time}
\end{center}
\end{figure}

We increased the original scanning speed of the observations from $2\fdg5$/min to $4\degr$/min after tests performed in 2006. 
The reason was to minimize the influence of system instabilities and also the contamination by changing ground radiation and low-level RFI. To achieve the same S/N 
ratio we thus mapped the same region more often. The sensitivity is not unique throughout the entire survey section, since 
the coverages of $\ell$ and $b$ maps differ.  
The effective integration time of each map pixel is at least 2.6~sec for total intensity and 1.9~sec for polarization,
where the integration time of map pixel observed in ``narrow band mode'' were divided by a 
factor of 2 to be comparable with that of ``broad band mode''.
The effective integration time of each sub-region in Stokes $I$, $U$ and $Q$ for 600~MHz bandwidth is shown in Fig.~\ref{integration_time}. 

\begin{table}
   \caption{Observational parameters}
   \label{obspara}
  {\begin{tabular}{lll}\hline
                                  
       &            &  \\
  System Temperature           & 22~K $\rm T_{a}$   \\
  HPBW         & $9\farcm5$   \\
  Subscan separation, sampling             & 3$\arcmin$ \\
  Scan velocity                     & 2\fdg5/min or 4\degr/min  \\
  Scan direction                    & $\ell$ and $b$    \\
  Typical rms-noise for $I$      &0.5 - 0.7~mK $\rm T_{B}$   \\
  Typical rms-noise for $U$ and $Q$      & 0.3 - 0.4~mK $\rm T_{B}$    \\
  Typical rms-noise for $PI$        & 0.3 - 0.5~mK $\rm T_{B}$     \\
  Central frequency               & 4800~MHz or 4963~MHz   \\
  Bandwidth                         & 600~MHz or 295~MHz    \\
  aperture efficiency            & 62\%         \\
  beam efficiency                & 67\%          \\
  T$_{B}$/S                       & 0.164 K/Jy        \\
  Main Calibrator                 & 3C286       \\
  Flux density                    & 7.5~Jy \\
  Percentage Polarization         & 11.3\% \\
  Polarization Angle              & 33\degr \\
                    &              &            \\\hline
  \end{tabular}}
\end{table}

\subsection{Data reduction}

Data reduction was done following the standard procedures, which were already described in Paper~I. Subsequently the following steps are
applied: For total intensity $I$ maps, a baseline was appropriately fitted for each sub-scan, which implies that 
large-scale diffuse emission exceeding the length of the sub-scans of $8\degr$ or $10\degr$ is not preserved. In addition,
spiky pixels were removed, distorted sub-scan sections 
or entire sub-scans were set to dummy values or, for smooth regions, replaced by interpolation of the two neighboring sub-scans. 
The baselines of the Stokes $U$ and $Q$ maps were usually not fitted in order to preserve extended polarized emission, unless 
strong ground radiation at low elevations clearly contaminates the data. Baseline distortion effects along scanning direction were suppressed by the 
``unsharp masking'' method \citep{Sofue79}. Spiky pixels and bad sub-scans were corrected in the same way as the $I$ maps.
We noted that afterwards many maps show residual distortions visible as inclined stripes, which seem to be caused by RFI-sources of unknown origin.
They were well removed by rotating the map that the stripes align 
to rows or columns of the map to apply the ``unsharp masking'' procedure and then
rotating the map back. 

Total intensity and polarization calibration was based on 3C286 (see Table~1).
Instrumental polarization from strong sources was removed by the ``REBEAM'' procedure of the NOD2 package
as described in Paper~I. This method reduces the ringlike instrumental response in polarized intensity $PI$ by about 50\%, 
which leaves some residual instrumental response of the order of 1\%. Our sensitivity limit on average  
(3~$\times$ rms) is 1.2~mK $\rm T_{B}$ (or 7.3~mJy/beam). Only strong sources exceeding about 0.7~Jy will show 
weak polarization response of instrumental origin, which will, however, in most cases confuse with diffuse Galactic emission.

Compact sources of the NVSS catalogue \citep{Condon98} were used to check the position accuracy. Finally, all individually edited maps 
were combined by applying the ``PLAIT''-algorithm \citep{Emerson88}, where the Fourier transforms of the maps were added and the final map 
is obtained by an inverse Fourier transform. ``PLAIT'', in addition, is able to suppress remaining low-level scanning effects still visible in a few individual maps. 
 
\subsection{Absolute zero-level restoration for Stokes $U$ and $Q$  } 

We have observed scans of up to 10$\degr$ in length aiming to recover extended structures as large as possible. 
All maps have a relative zero-level by arbitrarily setting the edge values of each scan to zero. Total intensity maps 
 (Stokes $I$) always miss a positive 
temperature offset, while the offsets for Stokes $U$ and $Q$ maps may be positive or negative. Polarized intensity, $PI$, of unknown intensity originating 
from Faraday rotated diffuse emission in the interstellar medium may exist everywhere and is not related to total intensity. Thus the true zero-level of the observed $U$ and $Q$ 
maps remains unknown. Therefore $PI$ and the polarization angle, $PA$, as calculated from $U$ and $Q$, need to be corrected as well. 
We note that relative polarization zero-level setting may be done in different ways, e.g. setting the mean value of $U$ and $Q$ of each scan to 
zero \citep{Junkes87}. After we combined maps observed along $\ell$ direction with maps along $b$ direction, the edge areas of the final combined maps differ from zero.  

Since polarized components are vectors, a missing large-scale component may lead to a misinterpretation of the
 observed data \citep{Reich06}. 
This is in particular important for polarized emission resulting from Faraday rotation, which  clearly dominates the Galactic polarization maps  at $\lambda$6\ cm. In Paper~I, \citet{Sun07} already presented a solution for this problem 
by adopting the three-year K-band (22.8~GHz) polarization data from WMAP \citep{Page07}, which have a correct 
zero-level. Missing large-scale $U$ and $Q$ emission at $\lambda$6\ cm is restored by scaling the K-band data by 
a factor of $(4.8/22.8)^{-2.8}$, 
according to a temperature spectral index of $\beta = -2.8$. This procedure also assumes that the $RM$ of the diffuse emission is not significant. 

For this second much larger section of the $\lambda$6\ cm polarization survey we slightly modified the method applied in Paper~I
by taking meanwhile available additional information into account. We now use the five-year release of the WMAP observations  
\citep{Hinshaw09}. We calculated the spectral index distribution between the polarized emission at 1.4~GHz \citep{Wolleben06} and
the K-band data for the entire survey section, smoothed to a common angular resolution of 2$\degr$. We obtained a mean 
spectral index of $\beta = -2.92\pm0.25$. We note that this spectral index is largely biased by the dominating
polarized emission from the bright Fan-region, which is Faraday thin at 1.4~GHz. This, however, is likely not the
case for the Galactic plane emission at 1.4~GHz from large distances. Current estimates of the synchrotron total 
intensity spectrum quote very similar spectral values between 1.4~GHz and 23.8~GHz (see \citet{Dickinson09} for a recent discussion), which we expect to be valid for the extrapolation of Faraday thin diffuse large-scale polarized emission 
from 22.8~GHz to 4.8~GHz as well.

We compared the WMAP K-band (22.8~GHz) and Ka-band (33~GHz) polarization data \citep{Hinshaw09} at 2$\degr$ angular
resolution for common extended polarization structures in the present survey area.
Clearly, the vast majority of patchy, weak polarization features in the two WMAP maps were not correlated and thus do
not show patches of polarized emission. This in turn means that an extrapolation of the polarized K-band emission 
towards $\lambda$6\ cm becomes questionable as it might introduce spurious features specific to the K-band map. 
Large-scale polarization gradients, 
however, are common in the K-band and Ka-band maps. We therefore decided to convolve the $\lambda$6\ cm $U$ and $Q$ survey maps and
the corresponding K-band maps to 2$\degr$ angular resolution after having removed a few strong and compact polarized sources, The 
convolved maps were split into sections, scaled by a factor of $(4.8/22.8)^{-2.9}$ and the difference values in their corner areas 
were determined. These difference values were used to define correction hyper planes in $U$ and $Q$ for each $\lambda$6\ cm 
survey section and were applied to the data at their original resolution. In Table~2 we list the $U$ and $Q$ intensities of the Urumqi
observations and the corresponding scaled K-map values together with the resulting correction values.  The maximum error introduced 
by assuming a constant spectral index of $\beta = -2.9$ will occur at $\ell = 129\degr$ and is estimated to be about $\pm$1.5~mK $\rm T_{B}$ 
in case the assumed spectral index varies by $\Delta\beta = \pm$0.1.

A significant $RM$ will change the extrapolated corrections for $U$ and $Q$, while $PI$ remains unchanged. Numerous $RM$s from 
extragalactic sources in the Galactic plane are available \citep{Brown07}. On average high $RM$-values are observed in the surveyed area with 
a clear gradient along $\ell$, but also a significant scatter of $RM$ is noted. However, it is known that the $RM$ of diffuse polarized Galactic emission 
in this area is small \citep{Spoelstra84,Haverkorn032}. We used the recent 3D-model of Galactic emissivities by \citet{Sun08}, which is in agreement 
with observed $RM$s, to model the polarized emission distribution at 4.8~GHz and 22.8~GHz. The distribution of $RM$ was obtained from the
$PA$ maps at both frequencies. The $RM$ map shows small values in general, as expected, and a nearly linear increase of $RM$ from $\ell = 129\degr$ to 
230$\degr$. For $\ell = 129\degr$, $b = +5\degr, 0\degr$ and $-5\degr$ the simulations predict $RM$ values of $-18$, $-35$ and $-25$~rad\ m$^{-2}$ . 
For $\ell = 230\degr$ the $RM$ values are $+31$, $+66$ and $+38$~rad\ m$^{-2}$, respectively. The simulations by \citet{Sun08} did not take into account the 
excessive polarized emission from the so-called "Fan"-region, a discrete very extended and highly polarized structure, which clearly dominates the large-scale 
polarized emission for $\ell$ lower than about $160\degr$. The ``Fan''-region is known to have $RM$s very close to zero at low angular resolution \citep{Spoelstra84}. 
Thus no $RM$ based correction for the low $\ell$ end of the observed area is necessary. The maximum $RM$ of 66~rad\ m$^{-2}$ at 
$\ell = 230\degr, b = 0\degr$ means a $PA$ change between 4.8~GHz and 22.8~GHz of 14$\degr$. Such an angle difference 
should be taken into account. However, the zero-level corrections for $U$ and $Q$ in this area (see Table~2) are below the 
3 $\times$ rms-noise in $U$ and $Q$ of the observations, so that we neglect the $RM$ effect on the $U$ and $Q$ corrections.

\begin{figure}
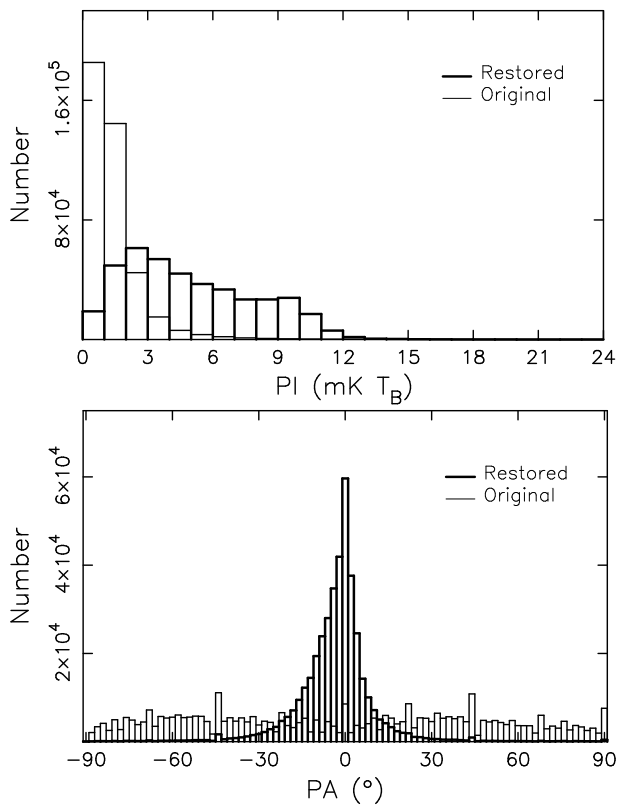

\begin{center}
\includegraphics[angle=-90,width=8cm]{13793fig2a.ps}
\includegraphics[angle=-90,width=8cm]{13793fig2b.ps}
\caption{Pixel distribution for $PI$ (top) and $PA$ (bottom) before and after restoring missing large-scale structures
 extrapolated from WMAP K-band. Details are discussed 
in Sect.~2.3.}
\label{restoration}
\end{center}
\end{figure}

The effect of the zero-level restoration process is illustrated in Fig.~\ref{restoration} by comparing the distributions of $PI$ and
$PA$ before and after the large-scale correction. Both distributions are clearly changed. In particular, $PA$ changed from an almost
uniform distribution between $-90\degr$ and $+90\degr$ into a distribution with a clear maximum for
 $PA$ near 0$\degr$ for the restored data, reflecting the fact that significant large-scale corrections are 
required for Stokes $Q$, while $U$ remains almost unchanged. This means that on large scales the magnetic field is orientated along $\ell$ (PA = 0$\degr$).\footnote {$PA$ is defined as the angle between E-vector and the Galactic north, which is equivalent to the angle between the B-vector and the Galactic plane. $PA$ runs counter-clockwise.}

\begin{table}
\caption{Hyper plane corrections in mK $\rm T_{B}$}
\label{sources}
\begin{tabular}{ccrrrrrr}
\hline\hline
   $\ell$ & $b$                   & U           & U            &U             & Q            & Q          & Q\\
 (deg)   & (deg)            & 6~cm    & Kmap      &corr          & 6~cm      & Kmap    & corr  \\
\hline
$230.0$    &$+5.0$     &$0.0$      &$0.0$       &$0.0$       &$-0.1$      &$0.5$     &$0.6$  \\        
$210.5$    &$+5.0$     &$-0.1$     &$-0.5$      &$-0.4$      &$-0.1$      &$0.7$     &$0.8$  \\
$188.0$    &$+5.0$     &$-0.6$     &$-0.4$      &$0.2$       &$0.2$       &$1.8$     &$1.6$  \\
$167.5$    &$+5.0$     &$-0.3$     &$-1.5$      &$-1.2$      &$0.6$       &$4.1$     &$3.5$  \\
$148.5$    &$+5.0$     &$-0.2$     &$-0.3$      &$-0.1$      &$0.2$       &$8.2$     &$8.0$  \\
$129.0$    &$+5.0$     &$0.5$      &$0.6$       &$0.1$       &$0.6$       &$10.1$   &$9.5$  \\
                      &                     &               &                 &                 &               &               &  \\
$230.0$    &$-5.0$      &$-0.5$      &$-0.6$      &$-0.1$      &$0.0$      &$2.3$      &$2.3$  \\
$210.5$    &$-5.0$      &$-0.2$      &$-0.5$      &$-0.3$      &$0.0$      &$3.0$      &$3.0$  \\
$188.0$    &$-5.0$      &$-0.2$      &$-1.3$      &$-1.1$      &$0.1$      &$5.4$      &$5.3$  \\
$167.5$    &$-5.0$      &$0.1$       &$-1.0$      &$-1.1$      &$-0.1$     &$5.7$      &$5.8$  \\
$148.5$    &$-5.0$      &$-0.1$      &$0.7$       &$0.8$       &$-0.6$     &$8.5$      &$9.1$  \\
$129.0$    &$-5.0$      &$-0.1$      &$0.3$       &$0.4$       &$-0.4$     &$10.2$    &$10.6$ \\
\hline
\end{tabular}\\
\end{table}

\section{Analysis of the survey region}

Large-scale emission seen in the present $\lambda$6\ cm survey section originates from the local arm, the Perseus arm and probably an 
outer arm \citep{Hou09}. Also emission contributed from the inter-arm regions is possible. A large number of extended and compact sources are 
also visible in the surveyed area. The total intensity $I$ maps very well resemble the emission structures visible in surveys at longer wavelengths 
with similar angular resolution, e.g. the Effelsberg surveys at $\lambda$21\ cm \citep{Kallas80,Reich97} and at $\lambda$11\ cm \citep{Fuerst90}. 
The $\lambda$6\ cm polarization data, however, show rather different structures compared to partly available Effelsberg $\lambda$21\ cm 
survey data \citep{Uyaniker99,Reich04} and are therefore of particular interest, so that we focus on them in the following. 

Compact or slightly resolved sources of the entire $\lambda$6\ cm Urumqi survey will be listed in a separate paper after completion
of the survey. Several prominent sources like the Cygnus Loop \citep{Sun06}, OA184 \citep{Foster06}, the SNRs G156.2+5.7 \citep{Xu07}, 
S147 \citep{Xiao08}, HB3 \citep{Shi08b}, and  G65.2+5.7 \citep{Xiao09} were already studied in detail based on observations made with the
Urumqi $\lambda$6\ cm system, which prove the high quality of the data. 
Data of three newly identified \ion{H}{II} regions from the present survey region were published
by \citet{Shi08a}. We will present a discussion of other SNRs, \ion{H}{II}-regions and prominent extended emission complexes in subsequent papers.

\subsection{The survey maps}

\begin{figure*}
\begin{center}
\includegraphics[width=0.9\textwidth]{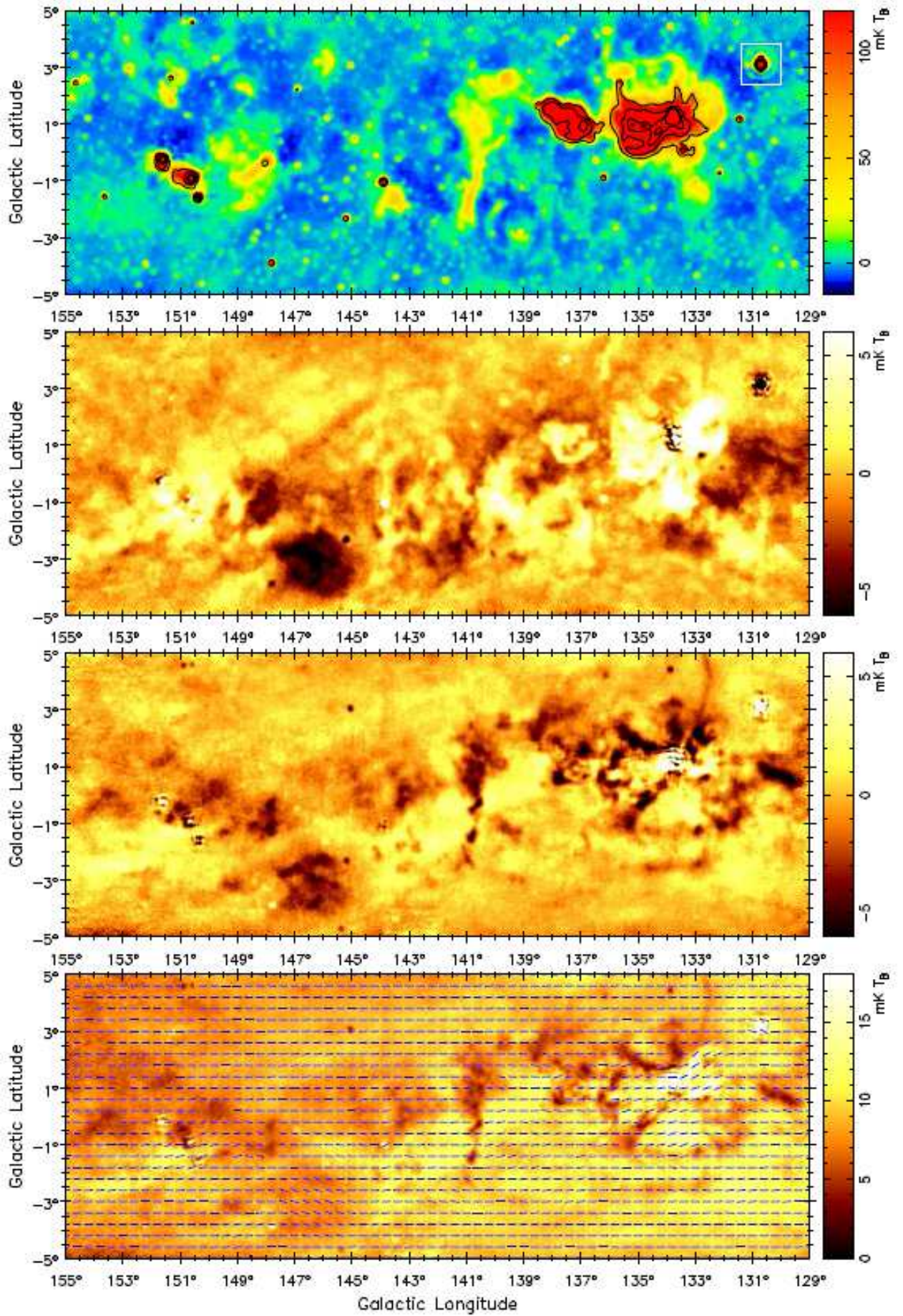}
\caption{Urumqi $\lambda$6\ cm survey maps (Part~1). From top to bottom: Stokes $I$, $U$, $Q$ maps in mK\ $\rm T_{B}$ as observed and 
the $PI$ map with large-scale restoration. $I$ contours are shown at 70, 200, 500 and 800~mK\ $\rm T_{B}$. 
The $PI$ map is overlaid by polarization vectors in B-field direction, when $PI$ exceeds 6~mK\ $\rm T_{B}$.
Vectors are shown for pixel being 24$\arcmin$ apart in $\ell$ and $b$. Their length is proportional to $PI$. The white rectangle marks the area of SNR 3C58, which is too strong 
to remove its side-lobes completely.}
\label{section1}
\end{center}
\end{figure*}

\begin{figure*}
\begin{center}
\includegraphics[width=0.9\textwidth]{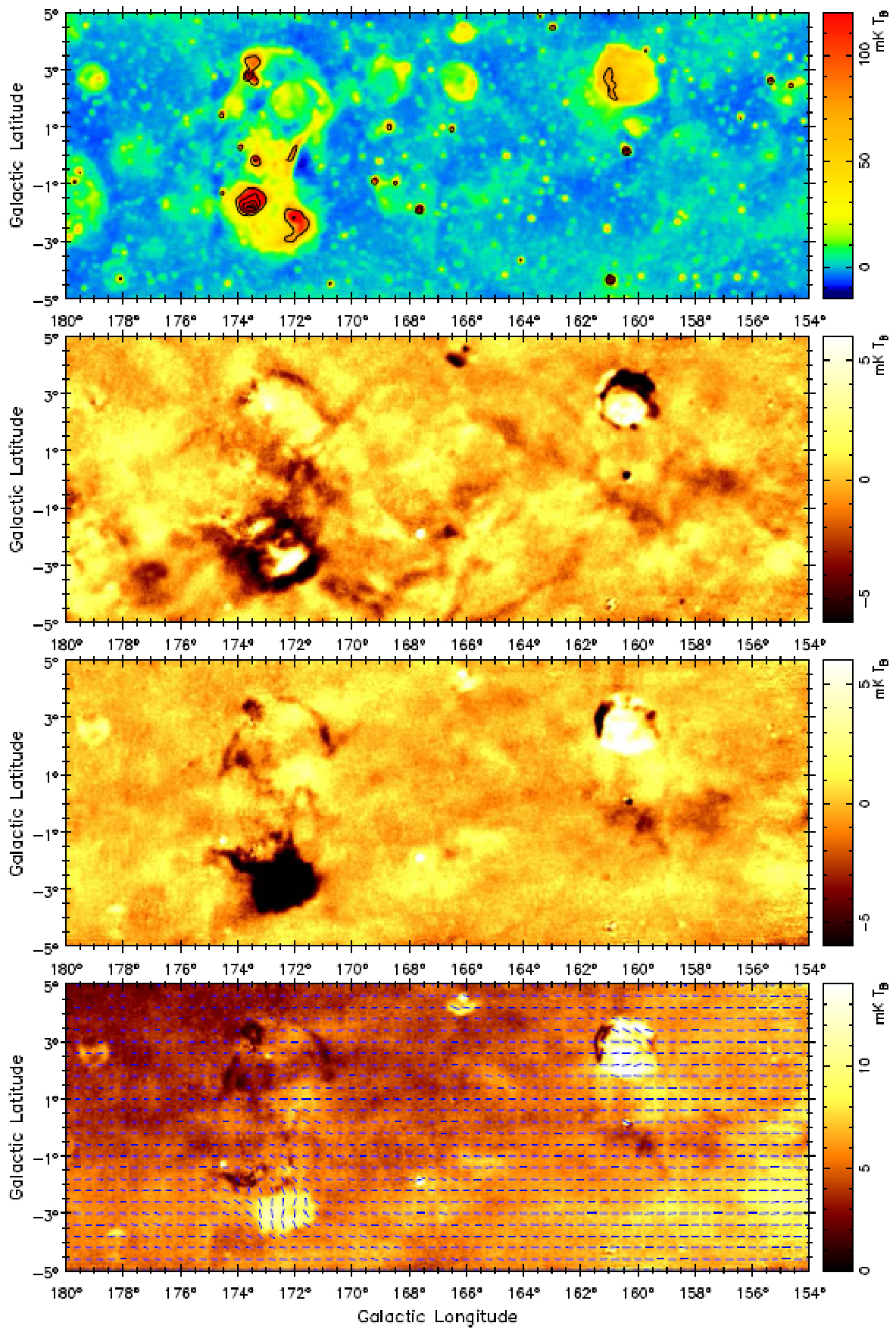}
\caption{Sequence and contours as in Fig.~\ref{section1} for Part~2. 
Vectors are shown for $PI$ exceeding 2.5~mK\ $\rm T_{B}$.}
\label{section2}
\end{center}
\end{figure*}

\begin{figure*}
\begin{center}
\includegraphics[width=0.9\textwidth]{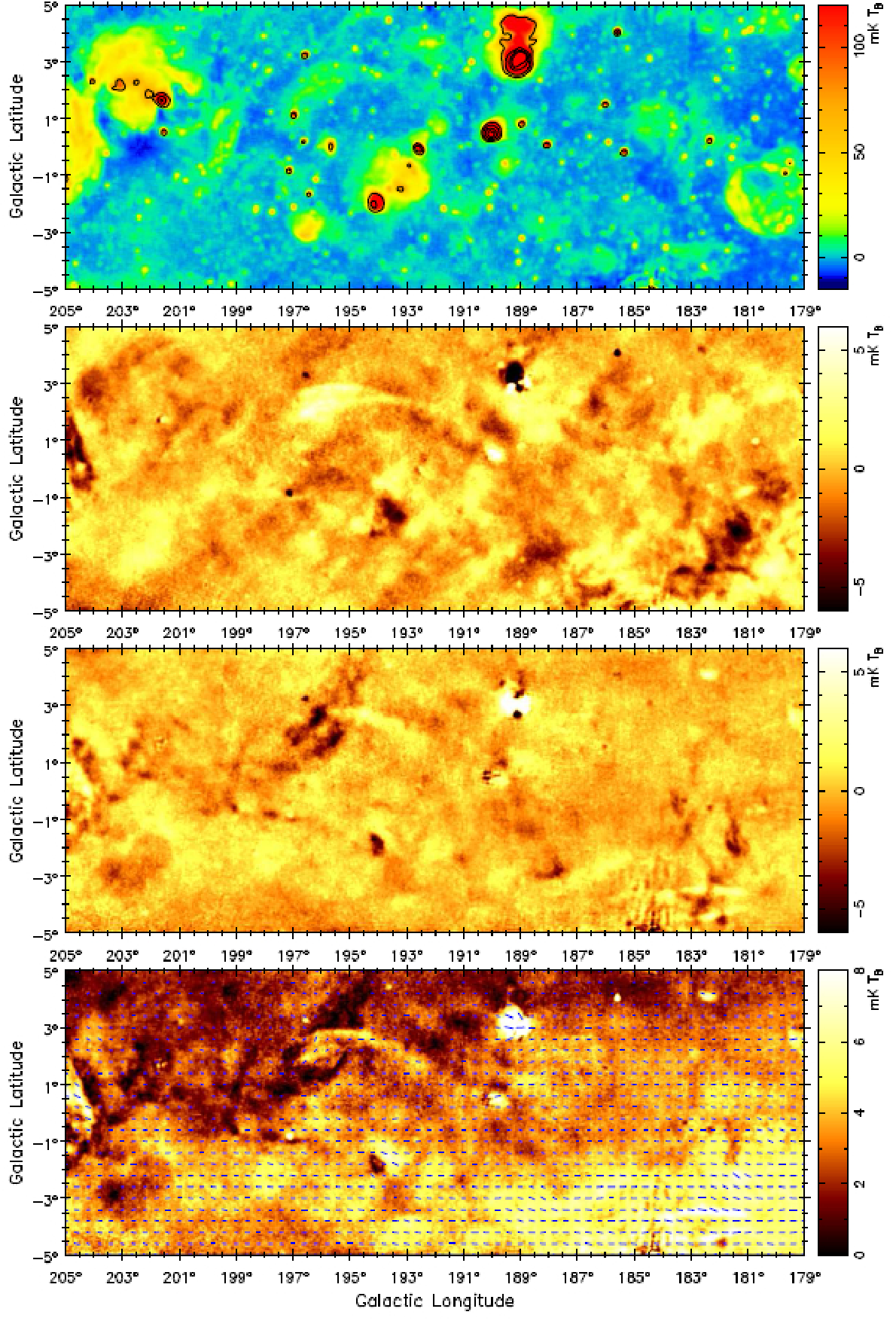}
\caption{Sequence and contours as in Fig.~\ref{section1} for Part~3.
Vectors are shown for $PI$ exceeding 1.5~mK\ $\rm T_{B}$.}
\label{section3}
\end{center}
\end{figure*}

\begin{figure*}
\begin{center}
\includegraphics[width=0.9\textwidth]{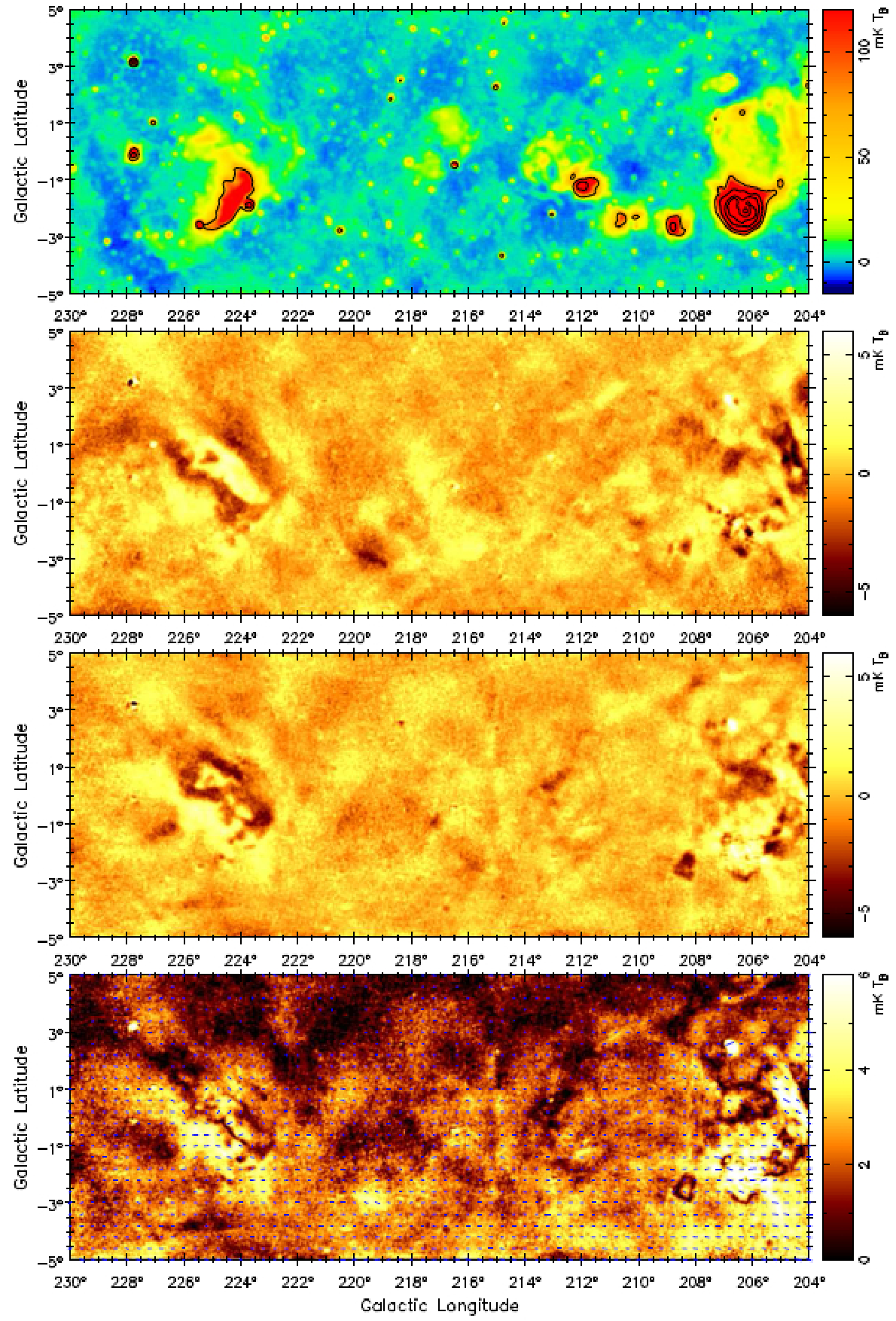}
\caption{Sequence and contours as in Fig.~ \ref{section1} for Part~4.
Vectors are shown for $PI$ exceeding 0.9~mK\ $\rm T_{B}$.}
\label{section4}
\end{center}
\end{figure*}

We show the maps of the outer Galactic plane in four parts (Part~1 to Part~4) in Figs.~\ref{section1} to \ref{section4}.
The maps have an overlap in $\ell$ of $1\degr$. We show Stokes $I$, $U$ and $Q$ maps as observed with the Urumqi 25-m telescope. 
In addition, we show maps of $PI$, which were calculated from the $U$ and $Q$ maps but with the large-scale hyper plane corrections as 
discussed in Sect.~2.3. When calculating $PI$ the correction for the positive noise offset: $PI^{2} = U^{2} + Q^{2} - 1.2 \sigma_{U,Q}^2$ \citep{Wardle74} 
was applied, where $\sigma_{U,Q}$ (see Table~1) is the averaged rms-noise for the $U$ and $Q$ maps for a specific survey area. 
Polarization bars in B-field direction are overlaid on the $PI$ image. The bars indicate the magnetic field direction in case of small Faraday rotation.
The sensitivity throughout the surveyed region varies slightly due to different integration time for different parts as outlined in Sect.~2.1 and Fig.~\ref{integration_time}. In addition  
smaller $\ell$ sections have in general a higher sensitivity than areas with larger $\ell$, because they benefit from both the ``broad band mode'' 
and less contamination by ground radiation.

The survey maps and the compact source list will be made publicly available after completion of the
entire project via the MPIfR survey-sampler\footnote{http://www.mpifr-bonn.mpg.de/survey.html}, the webpage at the
National Astronomical Observatories, CAS\footnote{http://www.nao.cas.cn/zmtt/6cm/} and possibly other 
data centres. 

\subsection{Supernova Remnants (SNRs)}

SNRs play an important role for many processes in the interstellar medium, such as energy input, chemical enrichment of heavy elements, cosmic ray production, 
and thus influence the evolution of galaxies. Most Galactic SNRs are well studied at low radio frequencies, while information about their 
fainter high-frequency emission is limited. The polarization properties of many SNRs are not well studied at all. Sensitive observations of large SNRs are quite time 
consuming for large single-dish telescopes at high frequencies because of their small beam-size and the low intensity of SNRs.  Observations with interferometers have even 
higher angular resolution but suffer from missing large-scale components. Our $\lambda$6\ cm survey complements high-frequency total power and polarization data for numerous large diameter SNRs.

Eleven known SNRs according to the most recent SNR catalogue\footnote{http://www.mrao.cam.ac.uk/surveys/snrs/}  \citep{Green09} are all visible in the present survey region. Several individual studies of SNRs
based on the Urumqi $\lambda$6\ cm survey were already published as mentioned above, more are in preparation. 
The sources HB3, OA184, S147 and the small 
bottom part of G156.2+5.7 are included in the present survey section. For the first time polarized 
emission at $\lambda$6\ cm is seen for the SNRs HB9 (G160.9+2.6), VRO42.05.01 (G166.0+4.3), the
Monoceros Nebula (G205.5+0.9), and PKS0646+06 (G206.9+2.3). 

So far we have not unambiguously detected any new SNRs. The surface brightness limit for a 
SNR with a thick shell according to a 3$\sigma$ detection limit is about $\rm \Sigma_{1GHz} = 3.9\ \times 10^{-23} [Wm^{-2}\ Hz^{-1}\ sr^{-1}$] 
for a temperature spectral index of $\beta = -2.5$. This is lower than that of G156.2+5.7 \citep{Reich92}, which has the lowest surface 
brightness in the SNR catalogues.

As an example to show the potential of the Urumqi $\lambda$6\ cm survey for SNR research, we
discuss the SNR candidate G151.2+2.85, which 
was proposed by \citet{Kerton07} based on a steep temperature spectrum ($\beta = -2.75$) of the filamentary structures designated CGPSE~172 and 
168 (see their Fig.~5), which are clearly visible at 408~MHz and 1420~MHz in the CGPS \citep{Taylor03}. 
These two filamentary structures are aligned and about 1$\degr$ long in total. Both filaments are seen in the 
$\lambda$6\ cm survey and 
the Effelsberg survey at $\lambda$11\ cm \citep{Fuerst90} and $\lambda$21\ cm \citep{Reich97}. From a TT-plot of $\lambda$6\ cm versus 
$\lambda$21\ cm data we obtained temperature spectral indices of $\beta = -2.44\pm0.08$ and $\beta = -2.34\pm0.08$ for CGPSE~172 and 168, 
respectively, larger than those of \citet{Kerton07}. These results clearly confirm the non-thermal nature of these filaments 
and support the suggestion by \citet{Kerton07} that the filaments are part of a SNR shell. However, like \citet{Kerton07} 
we can not give the entire 
size and integrated flux density of the SNR shell, because outside the filaments any SNR related emission is too faint, so that it confuses with unrelated Galactic emission. 
Additional observations, in particular outside of the radio range, are required to trace this object.

\subsection{\ion{H}{II} regions}

Despite of numerous \ion{H}{II} region catalogues, in particular that compiled by \citet{Paladini03}, there 
are many more
\ion{H}{II} regions to be detected and their physical parameters to be determined. 
In this survey region many large \ion{H}{II} regions have been detected.
At $\lambda$6\ cm the non-thermal to thermal emission
ratio is lower than that at longer wavelengths, so that the detection or isolation of \ion{H}{II} regions from diffuse Galactic non-thermal emission including survey data at longer wavelengths in the analysis can be done more easily. 
We have carefully analyzed the maps and searched for sources with flat spectra and strong infrared emission and found several compact and extended \ion{H}{II} regions,
previously not catalogued. We also noted that many \ion{H}{II} regions have not well defined parameters, which we are able to improve based on the new radio data. The results of this analysis will be presented in forthcoming paper.

\subsection{Prominent Faraday Screens} 

Faraday Screens are magnetized interstellar objects, which do not emit synchrotron emission themselves, but contain 
a regular magnetic field and thermal electrons
causing Faraday rotation. Depending on their physical parameters, Faraday Screens depolarize and rotate polarized 
background emission, which is observed 
together with the polarized foreground emission. The observed polarized emission surrounding a Faraday Screen may be either higher or lower than that seen in the Faraday Screen 
direction, depending on its $RM$ and on the properties of the foreground and background components. Faraday Screens become visible as coherent structures in 
$PI$ and/or $PA$ maps compared to the diffuse polarized Galactic emission. A proper analysis of Faraday Screens requires the inclusion
of polarized structures on all scales. Faraday Screens were already discussed and analyzed in various earlier papers e.g. \citet{Wolleben04}, \citet{Reich06}, {\and} Paper~I. 
Of particular 
interest are Faraday Screens detected in the $\lambda$6\ cm polarization survey maps, since they have larger $RM$s than those at $\lambda$11\ cm or $\lambda$21\ cm,
which might indicate regular magnetic fields with significant strength depending on their thermal electron densities and sizes.
If $PA$ is rotated by 180$\degr$ by a Faraday Screen the background remains unchanged except for beam depolarization. This corresponds to a $RM$ exceeding about 
70~rad\ m$^{-2}$ at $\lambda$21\ cm or 260~rad\ m$^{-2}$ at $\lambda$11\ cm, but about 800~rad m$^{-2}$ at $\lambda$6\ cm. The visibility of Faraday Screens in total intensity just depends on their thermal electron density. 
Thus we do not distinguish between \ion{H}{II} regions and Faraday Screens with no counterpart in total intensity or H$\alpha$ in the following.

\subsubsection{Modeling Faraday Screens}

A simple model to derive the physical parameters of a Faraday Screen was already introduced in Paper~I. The model uses two observed components, marked as 
``$\mathrm{on}$'' and ``$\mathrm{off}$'', where ``$\mathrm{on}$'' is for the position where the line of sight passes the Faraday Screen. Through fitting the Faraday Screen parameters 
by the observed data, the $RM$ and the depolarization properties of a Faraday Screen can be derived. 
The polarized background emission, $PI_{bg}$, is the component 
originating at larger distances than the Faraday Screen, and the polarized foreground emission, $PI_{fg}$, originates in front of the Faraday Screen. The observed ``$\mathrm{off}$'' 
component is simply the combined polarized background and foreground emission, while the ``$\mathrm{on}$'' component is the polarized foreground emission 
plus the modulated polarized background emission by the Faraday Screen. We assume both components 
are smooth and have the same $PA$
for the ``$\mathrm{on}$'' and ``$\mathrm{off}$'' components. The equations below describe the model, 
details can be found in Paper~I.

\begin{equation}
\displaystyle{
\left\{
\begin{array}{cc}
\displaystyle
\frac{PI_{\mathrm{on}}}{PI_{\mathrm{off}}}=\sqrt{\mathit{f}^2(1-c)^2+c^2+2\mathit{f}c(1-c)\cos2\psi_s}\ , \\ \displaystyle
PA_{\mathrm{on}} - PA_{\mathrm{off}}=\frac{1}{2}\arctan\left(\frac{\mathit{f}(1-c)\sin2\psi_s}{c+\mathit{f}(1-c)\cos2\psi_s}\right),
 &
\end{array}
\right.
}
\end{equation}
here $\mathit{f}$ is the depolarization factor ranging from 0 to 1, where $\mathit{f}$ = 0 indicates total depolarization by 
the Faraday Screen, while $\mathit{f}$ = 1 means no depolarization. Parameter $c$ is defined as $PI_{fg}/(PI_{fg}+PI_{bg})$.

Note that the model assumes that the $PA$s of the background and foreground emission are in general the same, because the dominating large-scale magnetic 
field is oriented along the Galactic plane. A model which takes into account different $PA$s for foreground and background emission was presented 
by \citet{Wolleben04}, which, however, at least needs observations at two wavelengths. The equations above show the dependence of the observed 
$PI_{\mathrm{on}}$ and $PI_{\mathrm{off}}$, the $PA_{\mathrm{on}}$ and the $PA_{\mathrm{off}}$ from the modeled foreground $PI_{fg}$ and 
background $PI_{bg}$ emission components and the Faraday rotation angle $\psi_{s} = RM_{FS} \lambda^{2}$.

Here, $RM_{FS}$~[rad~m$^{-2}$] = 0.81\ $n_{e}$[cm$^{-3}]\ B_{\parallel}[\mu$G]\ $l$[pc], with $n_{e}$ being the thermal electron density, 
$B_{\parallel}$ the field strength of the line-of-sight component of the regular magnetic field and $l$ the line-of-sight length of the Faraday Screen. The size of the source 
is usually assumed to be that seen in projection. In case the Faraday Screen has measurable thermal emission, $n_{e}$ can be calculated from the emission measure $(EM)$, 
which is defined as $EM=n_{e}^2\ l$. For an optically thin \ion{H}{II} region, the observed brightness temperature $T_{B}$ depends on $EM$,

\begin{equation}\label{brightness temperature}
T_{B}=8.235\times10^{-2}\left(\frac{T_e}{K}\right)^{-0.35}\left(\frac{\nu}{GHz}\right)^{-2.1}\left( \frac{EM}{pc\ cm^{-6}}\right)a(\nu,T)\ ,
\end{equation}
where the correction a($\nu$, T) is taken as 1. The effective temperature $T_{e}$ is always assumed to be 8000~K. Another method to obtain $n_{e}$ 
depends on the measured H$\alpha$ intensity, which needs to be corrected by the reddening measurement of the exciting star following \citet{Haffner98}:

\begin{equation}
EM=2.75\ T_{4}^{0.9}I_{\rm H\alpha}\exp\ [2.44E(B-V)]\ .
\end{equation}
Precise reddening measurements are difficult to obtain.

Prominent extended Faraday Screens seen in this survey section were selected for discussion in the following 
in order of their $\ell$.

\subsubsection{W5 ($\ell = 137\fdg6, b = 1\fdg1$) and the ``lens'' Faraday Screen}

\begin{figure}
\begin{center}
\includegraphics[width=0.36\textwidth]{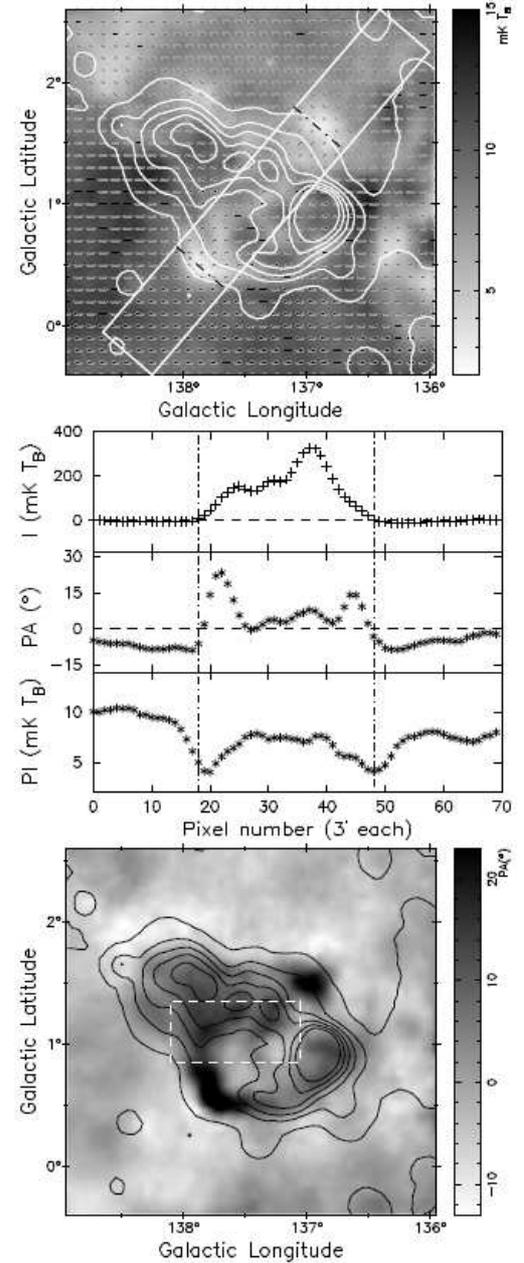}
\caption{Upper plot shows the $PI$ map of W5 in grey scale, overlaid by contours of $I$. The contour lines run from the local 3$\sigma$ level of 4.5~mK\ $\rm T_{B}$  in steps of 80~mK\ $\rm T_{B}$. The bars indicate B-vectors with their length proportional to $PI$ and are shown for every second pixel ($6\arcmin$) in $\ell$ and $b$ direction. The middle plot shows average values of $I$, $PA$ and $PI$ for the inclined rectangular
region marked by the solid line in the upper plot. The dashed-dotted line in the upper and middle plots mark the boundary 
of W5. The lower plot shows the $PA$ distribution in grey scale. The rectangular region marks the ``lens'' structure visible at $\lambda$21\ cm \citep{Gray98}. 
}
\label{w5}
\end{center}
\end{figure}

W3/W4/W5 are prominent \ion{H}{II} regions forming a chain together with the SNR~HB3 in the Perseus~arm about 2~kpc away.
\citet{Gray99} used the DRAO Synthesis Telescope and obtained 1$\arcmin$ resolution radio images of both total intensity and 
polarized emission of this field at 1.4~GHz. We limit our discussion to W5 in the following. At $\lambda$6\ cm $PI$ in this area (Fig.~\ref{w5}) 
appears to be depolarized by a different amount and the distribution is mottled for our $9\farcm5$ beam, although we see less fine structures when 
compared to the 1.4~GHz map of \citet{Gray99}.

\citet{Heiles00} compiled a catalogue of polarized stars and lists nine of them in the 
vicinity of W5. We calculated a mean $PA$ = 0\fdg3$\pm4\fdg1$ up to the largest distance of 3.6~kpc, 
where $PA$ runs counter-clockwise with the Galactic plane as reference.
The alignment of $PA$ around 0$\degr$ means that the magnetic field is orientated along the
Galactic plane for all distances and thus confirms the assumption of our Faraday Screen-model. 
The $\lambda$6\ cm $PA$s in the W5 area, however, vary (Fig.~\ref{w5}) and mottled depolarization is 
also visible, which is explained by Faraday rotation of different amount.

\citet{Westerhout58} listed some physical parameters of W5 such as EM = 4000~pc\ cm$^{-6}$ and $n_{e} = 10\ cm^{-3}$.
From the $\lambda$6\ cm brightness temperature of W5 West of about 380~mK\ $\rm T_{B}$, we calculated 
a comparable $EM$ value of 2900~pc\ cm$^{-6}$ according to Eq.~2. In addition the H$\alpha$ intensity of W5 can be extracted 
from the H$\alpha$ full sky map \citep{Finkbeiner03} to be about 300~Rayleigh. A reddening measurement of the exciting star 
$BD+59\fdg0562$, also named Hilt~360 \citep{Hiltner56}, gives an  E(B-V) factor of 0.63. Following Eq.~3, $EM$ can be estimated 
to be 3140~pc\ cm$^{-6}$, slightly above the radio-based result.

If we take $|B_{\parallel}| = 3\ \mu$G as assumed by \citet{Gray98} as the strength of the line-of-sight component of the regular 
magnetic field within W5, the expected $|RM_{FS}|$ value is about 970~rad\ m$^{-2}$ for a size of 40~pc and for $n_{e}$ about 
10\ cm$^{-3}$  \citep{Westerhout58}. This predicts about 217$\degr$ for $\psi_{s}$ at $\lambda$6\ cm. 

W5 is a large object, where the foreground polarization fraction $c$ may vary across the source.
For $c = 0.7, 0.8$, and $0.9$, and assuming $B_{\parallel}$ and $n_{e}$ to be uniform, we calculate $PI_{\mathrm{on}}/PI_{\mathrm{off}} = 84\%, 88\%, 93\%$ 
and $PA_{\mathrm{on}} - PA_{\mathrm{off}} = 10\degr, 6\degr, 3\degr$ for the Faraday Screen.
As an example, we model $RM_{FS}$ for the fairly uniform W5 East area for an average brightness temperature of about 270~mK\ $\rm T_{B}$ and a
size of $0\fdg7$ corresponding to 24~pc for 2~kpc distance. With $c = 0.79$ we calculate $RM_{FS}$ = 260$\pm$60 rad\ m$^{-2}$, which is 
much smaller than the value estimated for W5 above. This indicates that $B_{\parallel}$ is about 1.5\ $\mu$G and $n_{e}$ about 9.2\ cm$^{-3}$ within W5 East.

\begin{figure}
\begin{center}
\includegraphics[angle=-90, width=0.39\textwidth]{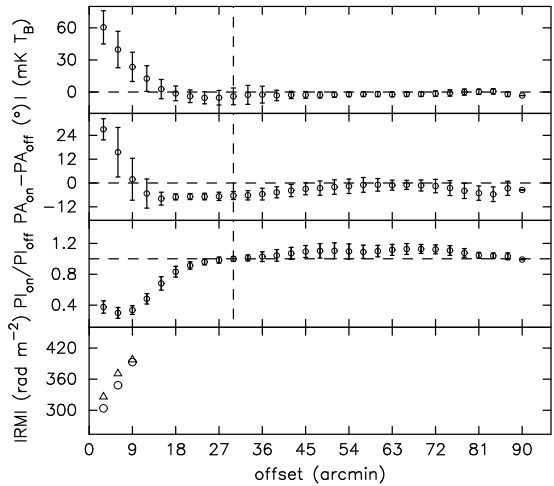}
\caption{Averaged observed values for the low-latitude W5 blob with centre coordinate ($\ell = 137\fdg80,b = 0\fdg54$) of a half
ring. $I$, $PA$ difference, $PI$ ratio and fitted $RM$ are shown from top to bottom, respectively. 
The vertical dashed line indicates the boundary between the ``$\mathrm{on}$'' and ``$\mathrm{off}$'' components.}
\label{w5_blob}
\end{center}
\end{figure}

Two remarkable polarization features are clearly visible at the edges of W5 in the $\lambda$6\ cm polarization map 
(Fig.~\ref{w5}) at $\ell =137\fdg85, b = 0\fdg50$ and at $\ell = 136\fdg85, b = 1\fdg60$,
which resemble Faraday Screens detected at the edges of molecular clouds by \citet{Wolleben04}.
We apply the model fit to the eastern blob. 
The problem is that we can not definitely decide from single-wavelength data whether $RM_{FS}$ is 
positive or negative, since the absolute values are very similar. We either obtain $RM_{FS} = -365\pm30$~rad\ m$^{-2}$
for $c = 0.30$ and  $\mathit{f} = 0.87$ or $RM_{FS} = +350\pm$40~rad\ m$^{-2}$ for $c = 0.60$ and $\mathit{f} = 0.93$.  We show the averaged observed values and the fitted $RM$s in Fig.~\ref{w5_blob}.
If the blob is a Perseus arm object like W5, a positive $RM_{FS}$ is preferred, because of the larger $c$ value.
\citet{Brown03b} examined the $RM$ values of extragalactic sources and pulsars in the direction of $105\degr \leq \ell \leq 135\degr$ within the Galactic plane and found that most values are negative. However, \citet{Mitra03} showed 
a schematic model (their Fig.~5) that the curvature of the magnetic field lines near \ion{H}{II} regions may
result in a reverse $RM$ sign. 
Assuming 2~kpc distance the first three pixels give an average $n_{e}$ of about 4.7~cm$^{-3}$ assuming a blob-size of
10.5~pc, and $B_{\parallel}$ is about 8.8\ $\mu$G. These parameters clearly differ from the average values obtained for W5.
Of course, we could not entirely rule out a projection effect, so that these blobs are seen at the periphery of W5 by chance.

\citet{Wolleben04} discovered line-of-sight magnetic field components exceeding 20\ $\mu$G at the surface of the local 
Taurus molecular clouds, which is morphologically quite similar to the W5 polarization blobs seen at $\lambda$6\ cm.
Note that the uncertainty of the line-of-sight size of the Faraday Screen plays an important role in determining  
$B_{\parallel}$ and $n_{e}$. A tube-like shaped Faraday Screen would reduce the values of $B_{\parallel}$ and $n_{e}$.  
To precisely constrain such high $RM$ values, observations at even shorter wavelengths than $\lambda$6\ cm  
are needed, which are, however, difficult to do.

An elliptical polarized ``lens'' structure was reported by \citet{Gray98} seen towards the central part of W5 at 1.4~GHz. 
This elliptical structure was also observed by \citet{Uyaniker04} at the same frequency using the Effelsberg 100-m telescope. 
Our $\lambda$6\ cm polarization data  (Fig.~\ref{w5}), however, does not show such kind of regular Faraday Screen feature. 
\citet{Gray98} quote a $\Delta$RM of 110~rad\ m$^{-2}$, which should have an effect at $\lambda$6\ cm. However, the 
new 1.4~GHz polarization survey by Landecker et al. (2010, submitted), which includes large-scale polarization information, reduces
the $RM$ attributed to the ``lens'' by about a factor of 10, which makes the ``lens'' almost invisible at $\lambda$6\ cm.

\subsubsection{The ``Drumstick'' at $\ell$ = 140$\degr$}

\begin{figure}
\begin{center}
\includegraphics[width=0.33\textwidth]{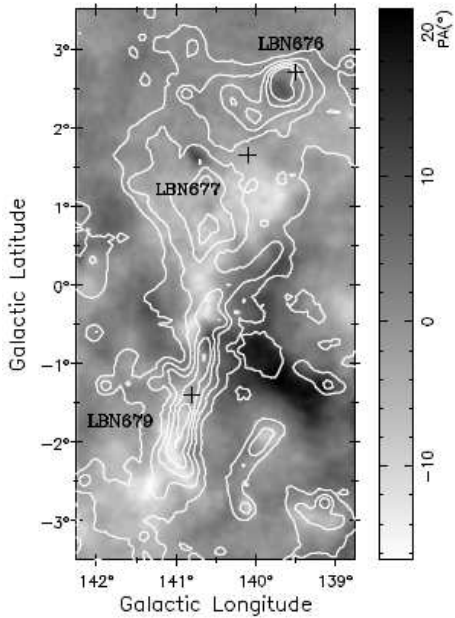}
\caption{$PA$ distribution in B-field direction for the ``drumstick'' area. $I$ contours start at 1.5~mK\ $\rm T_{B}$ and run in steps of 12~mK\ $\rm T_{B}$. Crosses show the central positions of the optically visible \ion{H}{II} regions.}
\label{drumstick}
\end{center}
\end{figure}

Several \ion{H}{II} regions are located around $\ell = 140\fdg5$ within a $3\fdg5 \times 6\fdg0$ field (Fig.~\ref{drumstick}). 
Unfortunately, no information of the \ion{H}{II} regions was given in the catalogue of \citet{Paladini03}. Three faint optically 
visible \ion{H}{II} regions of the Lynds catalogue are: the semi-ring shaped LBN~676 ($\ell = 139\fdg57, b = 2\fdg70$) with a 
size of $0\fdg8$, LBN~677 (SH 2-202) ($\ell = 140\fdg07, b = 1\fdg64$, size of $2\fdg0$), and the bar-like shaped LBN~679 
($\ell = 140\fdg77, b = -1.42\degr$, size of $2\fdg0$). For morphology reasons we name the three \ion{H}{II} regions the ``Drumstick'' 
 in the following.

LBN~676 and LBN~679 are supposed to be at the same distance in the Perseus arm. They were already investigated in some
detail by \citet{Green89} using DRAO Synthesis Telescope data at 408~MHz and discussed together with infrared and HI maps. \citet{Green89} found that
the elongated \ion{H}{II} region LBN~679 coincides with a large \ion{H}{I} spur located in the Perseus arm and pointed out that it 
is a thermal rather than a non-thermal feature as suggested by \citet{Kallas83}. 

\citet{Karr03} studied all three \ion{H}{II} regions with 1$\arcmin$ resolution using CGPS data at 1.4~GHz \citep{Taylor03}. The thermal character of 
LBN~679 was again confirmed and in addition they derived a thermal spectrum with $\beta = -1.85\pm0.1$ for LBN~676, thus excluding 
a possible SNR identification considered by \citet{Green89} for this shell structure. The $\lambda$6\ cm data agree with 
the thermal properties of all objects through a check of their temperature spectral indices via TT-plots using Effelsberg $\lambda$21\ cm data for comparison. 

All three LBNe appear to be depolarized at $\lambda$6\ cm when large-scale polarized emission is added. Estimates of the magnetic field strength from modeled 
$RM_{FS}$ require the thermal radio continuum brightness temperature to find their $EM$. For a known distance the source size and $n_{e}$ need to be 
calculated in addition. \citet{Green89} assumed that LBN~676 and LBN~679 are both at a distance of 3~kpc in the Perseus arm. However, the Perseus arm distance 
was revised by \citet{Xu06} to be about 2~kpc by triangulation of W3OH, which is just about 7$\degr$ apart in Galactic longitude. In the following we adopt this distance. 

\begin{figure}
\begin{center}
\includegraphics[angle=-90, width=0.37\textwidth]{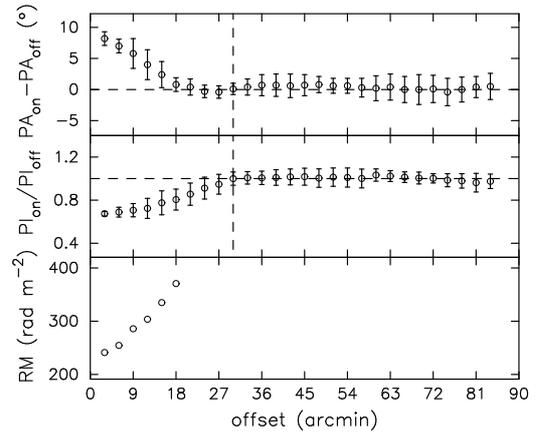}
\caption{Average values in northern direction for the \ion{H}{II} region LBN~676. Offsets are relative to the centre 
coordinate ($\ell = 139\fdg65, b = 2\fdg45$).
$PA$ difference and $PI$ ratio are shown in the upper and middle panel, respectively. The vertical dashed line indicates 
the boundary between the ``$\mathrm{on}$'' and ``$\mathrm{off}$'' components.}
\label{lbn676}
\end{center}
\end{figure}

It turns out that a ring average of the PA/PI differences for LBN~676 is difficult to perform because of confusion with LBN~677 emission. Thus we just take the data from 
the upper part. The model fit gives the best result (Fig.~\ref{lbn676}) for the first five pixels as $RM_{FS} = 280\pm$30~rad\ m$^{-2}$. Foreground polarized 
emission comprises about 79\% while $\mathit{f} = 0.80$. Likely most of the polarized emission originates in the local arm.
The total intensity attributed to LBN~676 is about 50~mK\ $\rm T_{B}$ at $\lambda$6\ cm. An apparent diameter of $0\fdg8$ 
equals to a path length of 28~pc for 2~kpc distance. We obtain $n_{e}$ as about 3.3\ cm$^{-3}$ and $B_{\parallel}$ $\approx$ +3.7\ $\mu$G. 

\begin{figure}
\begin{center}
\includegraphics[angle=-90, width=0.39\textwidth]{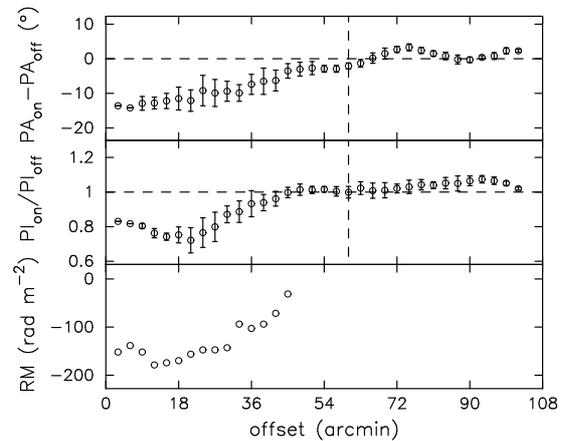}
\caption{Average values starting from the coordinate ($\ell = 141\fdg10, b = -2\fdg00$) to the outskirts of the \ion{H}{II} region LBN~679
in a sector from 10$\degr$ to 45$\degr$ counting from south (0$\degr$) to west .}
\label{lbn679}
\end{center}
\end{figure}

In the southern part of the 2$\degr$ long filamentary LBN~679, we find large $PA$ changes  and also in its outskirts beyond.
There is an inclined elongated $PA$ structure, about 1$\degr$ long, running from northeast to southwest. 
Model fitting (Fig.~\ref{lbn679}) is done for a 35$\degr$ wide cone in south-western direction of the rim. 
Best fit for the first ten pixels gives $RM_{FS} = -155\pm$15~rad\ m$^{-2}$, a foreground polarization fraction of 
about 60\% for $\mathit{f} = 0.85$. From the $\lambda$6\ cm brightness temperature 
of about 40~mK\ $\rm T_{B}$ and assuming a line-of-sight length of the source of 35~pc, we calculated 
$B_{\parallel}$ = $-2.0~\mu$G and $n_{e}$ = 2.7~cm$^{-3}$.
To the lower left direction of LBN~679, another $PA$ structure is located at about $\ell = 141\fdg35, b = -2\fdg60$. 

\begin{figure}
\begin{center}
\includegraphics[angle=-90, width=0.4\textwidth]{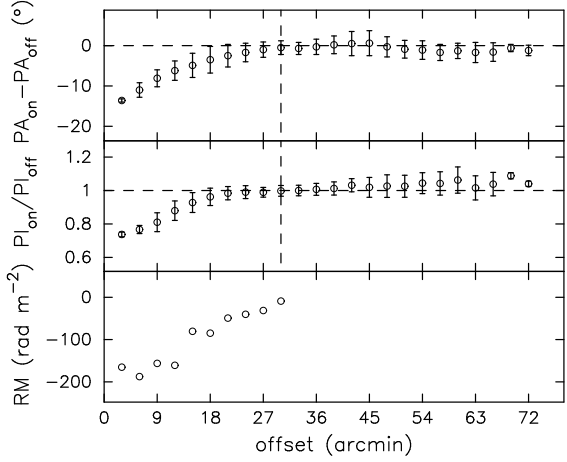}
\caption{Half ring average values from centre coordinate ($\ell = 139\fdg70, b = 1\fdg10$) towards the west of LBN~677.}
\label{lbn677}
\end{center}
\end{figure}

 \citet{Karr03} found that  the northern part of LBN~677 is associated with the exciting star HD~19820 at 1~kpc distance. 0.8~kpc distance
were quoted for associated CO-emission \citep{Blitz82}, which we adopt in the following discussion. 
A circular average is difficult to apply to the large area of LBN~677 due to significant variations and the non-symmetric 
distribution of $PA$. Different signs of $PA$ in different areas indicate changing properties throughout the entire region. 
We select the western part of LBN~677, where $PA$ changes smoothly for modeling. Based on the central five pixels average, the best fit (Fig.~\ref{lbn677}) 
gives a $RM_{FS}$ value of $-150\pm$40 rad\ m$^{-2}$. 69\% of the total polarized emission originates in the foreground. The depolarization factor is 
$\mathit{f}$ = 0.88. The 2$\degr$ diameter source gives a depth of 28~pc assuming a spherical shape. With the $\lambda$6\ cm brightness temperature of about 
20~mK\ $\rm T_{B}$ we calculate $EM$ = 137~pc\ cm$^{-6}$. $B_{\parallel}$ then is $-3.0~\mu$G and $n_{e}$ about 2.2~cm$^{-3}$.
The reddening measurement of the exciting star HD 19820 ($\ell = 140\fdg12, b = 1\fdg54$) is about 0.82 \citep{Hiltner56}. The H$\alpha$ intensity is about 
14~Rayleigh \citep{Finkbeiner03} after subtracting a background level of 11~Rayleigh.  We obtain $EM$ as 233~pc\ cm$^{-6}$ by Eq.~3.  
Then $B_{\parallel}$ reduces to $-2.3~\mu$G and $n_{e}$ increases to 2.9\ cm$^{-3}$. The physical parameters from both methods are similar. The best fit of 
$PI_{fg}$ for LBN~676 and LBN~679 is around 70\%. We found almost the same value also for LBN~677, which is unexpected in view of their different 
distances. If the polarized emissivity in this direction is fairly constant, this may indicate that LBN~677 is also at about 2~kpc distance.
For this distance $B_{\parallel}$ will change to $-1.9~\mu$G and $n_{e}$ to 1.4\ cm$^{-3}$.

We summarize the derived parameters for the three LBNe in Table~3. We note that our modeled $B_{\parallel}$ are rather similar to those found for 
\ion{H}{II} regions with a similar low electron density, e.g. S264 \citep{Heiles81} and G124.9+0.1 (Paper~I).

\begin{table}
\caption{Physical parameters for three LBNe}
\label{sources2}
\begin{tabular}{cccc}
\hline\hline\\
   \ion{H}{II} region                  &LBN~676               &LBN~677                &LBN~679\\
\hline\\
   $EM$(pc\ cm$^{-6}$)          &305                        &137 (233)                                  &255\\ 
   $n_{e}$(cm$^{-3}$)                                &3.3                        &1.4$^{a}$/2.2$^{b}$ (2.9$^{b}$)          &2.7\\
   $RM_{FS}$(rad\ m$^{-2}$)          &280                        &-150                                  &-155\\
   $B_{\parallel}$($\mu$G)                     &3.7                         &-1.9$^{a}$/-3.0$^{b}$ (-2.3$^{b}$)       &-2.0\\
\hline\\
\multicolumn{4}{l}{$^{a}$ is derived for a distance of 2~kpc;}\\
\multicolumn{4}{l}{$^{b}$ is derived for a distance of 800~pc.}\\
\end{tabular}\\
For LBN~677 we use the two $EM$ values derived from radio\\ 
emission and H$\alpha$ emission (see Sect.~3.4.3).\\
\end{table}

Besides the three LBN objects there are more $\lambda$6\ cm Faraday Screen structures in the field of Fig.~\ref{drumstick}. Some were listed below, 
which we, however, will not analyze in detail in this paper. A large area with very uniform $PA$s, which deviate about 22$\degr$ from the 
Galactic plane direction, can be identified in southwestern direction of the ``drumstick''. This Faraday Screen is centered at around 
$\ell = 139\fdg70, b = -1\fdg20$. There is no counterpart in the $I$ map (see Fig.~\ref{drumstick}), thus the thermal
electron density must be very low. Polarization observations at other wavelengths are needed to analyze
this feature in some detail. Also the small objects at $\ell = 139\fdg95, b = -2\fdg05$ and at $\ell = 139\fdg20, b = -3\fdg00$ were found to be thermal (Gao et al., in prep.) and act as Faraday Screens. They also can be clearly identified in the $PA$ map (Fig.~\ref{drumstick}).

\subsubsection{G146.4-3.0}

\begin{figure}
\begin{center}
\includegraphics[width=0.4\textwidth]{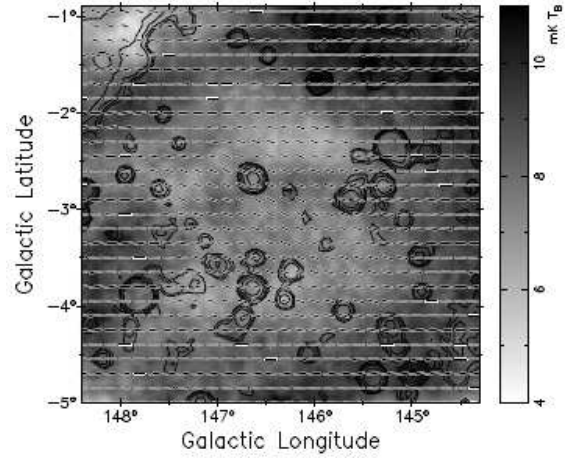}
\caption{$PI$ of the Faraday Screen G146.4-3.0 in grey scale. $I$ contours are overlaid running from $3\sigma$, or 1.5~mK $\rm T_{B}$, in steps of 2$^{n-1}\ \times$\ 1.0~mK $\rm T_{B}$.
Bars show B-vectors for every third pixel ($9\arcmin$) in $\ell$ and $b$ direction.}
\label{g146_fs}
\end{center}
\end{figure}

G146.4-3.0 is a large Faraday Screen with a diameter of approximately $3\fdg3$, centered at $\ell = 146\fdg40, b = -3\fdg00$. This structure shows highly polarized 
emission in the original $PI$ map at $\lambda$6\ cm and becomes depolarized after large-scale polarized emission is added. No counterpart can be found 
in the $I$ map (see Fig.~\ref{g146_fs}). The H$\alpha$ intensity in this region is in general low and does not show any excess related to this Faraday Screen.

\begin{figure}
\begin{center}
\includegraphics[angle=-90, width=0.4\textwidth]{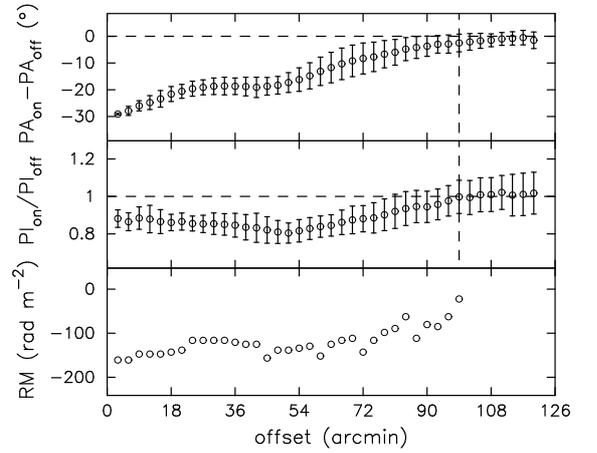}
\caption{Ring averaged data and modeled $RM$s for Faraday Screen G146.4-3.0 shown in Fig.~\ref{g146_fs}.}
\label{g146_fit}
\end{center}
\end{figure}

A Faraday Screen model fit (Fig.~\ref{g146_fit}) results in an average $RM_{FS}$ for the central area of 1$\degr$ diameter of $-140\pm$20~rad\ m$^{-2}$.  
26\% of total $PI$ in this area originates in front of the Faraday Screen for $\mathit{f}$ = 0.97. The $RM$ value is 
not exceptionally high, but the scatter across the Faraday Screen is large enough to make the object invisible in the $\lambda$21\ cm polarization 
survey by Landecker et al. (2010, submitted), although their survey only covers the northern area of the Faraday Screen. 

\begin{figure}
\begin{center}
\includegraphics[angle=-90, width=0.38\textwidth]{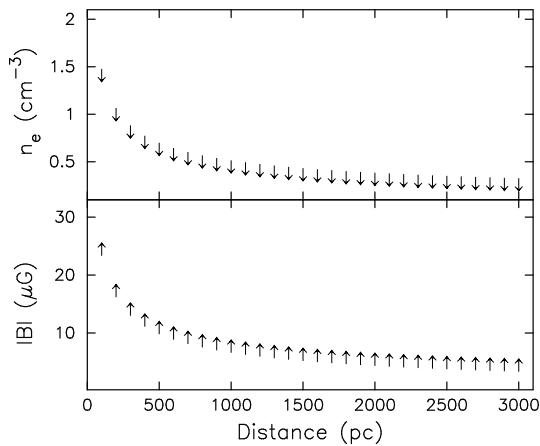}
\caption{Dependence of the central $B_{\parallel}$ and $n_{e}$ from the distance to the Faraday Screen G146.4-3.0.}
\label{g146_BNe}
\end{center}
\end{figure}

The distance to this prominent Faraday Screen in the $\lambda$6\ cm survey is not clear. To derive $n_{e}$ by its 
free-free $EM$ we used the 3$\sigma$ level of the $\lambda$6\ cm $I$ map, means about 1.5~mK\ $\rm T_{B}$, as a brightness temperature upper limit. 
The fitted central $B_{\parallel}$ and $n_{e}$ are estimated and shown in Fig.~\ref{g146_BNe} as a function of distance. 
Note that $n_{e}$ is always an upper limit, while  $B_{\parallel}$ has to be taken as minimum.

The foreground polarized emission of about 26\% is small compared to the fraction of about 70\%
obtained for the LBNe in the Perseus arm (see Sect.~3.4.3). Thus the distance to G146.4-3.0 should be small. 
Assuming the polarized emissivity along the line of sight to be the same as to the Perseus LBNe of about 3.4~mK\ kpc$^{-1}$, 
we estimate the distance to Faraday Screen G146.4-3.0 as about 690~pc, which implies a diameter of the Faraday Screen of about 40~pc. 
Then $B_{\parallel}$ for its central area is about $-8.6~\mu$G and $n_{e}$ is about 0.5\ cm$^{-3}$. 
However, the local emissivity is known to be two to three times larger than the average emissivity at 2~kpc distance \citep{Fleishman95}.
This will reduce the distance and the size of the Faraday Screen accordingly. For a distance of 280~pc, its size reduces 
to about 16~pc,
$n_{e}$ increases to about 0.8\ cm$^{-3}$ and $B_{\parallel}$ to $-13.5~\mu$G.

\citet{Taylor09} re-analyzed the NVSS \citep{Condon98} polarization data. $RM$s for 37\,543 polarized radio sources were 
derived. We examine our Faraday Screen model result by checking the $RM$s in the area of Fig.~\ref{g146_fs}. 
Two sources are located within the area of the Faraday Screen. The $RM$s of the two sources $\ell = 146\fdg62, b = -2\fdg69$ and 
$\ell = 146\fdg63, b = -3\fdg81$ are $-151.1\pm$8.2~rad\ m$^{-2}$ and $-221.9\pm$8.4~rad\ m$^{-2}$, respectively.
For another source at $\ell = 145\fdg61, b = -2\fdg9$ \citet{Brown03a} reported a $RM$ value of $-338\pm$15~rad\ m$^{-2}$ from the CGPS, which
is the most excessive $RM$ value. Outside the Faraday Screen area \citet{Taylor09} list $RM$s for 15 sources with an average $RM = -75.7\pm$30.4~rad\ m$^{-2}$.
Despite significant scatter the three sources show excessive negative $RM$s, which seem to be attributed to G146.4-3.0.

\subsubsection{Large magnetic bubbles at $\ell = 165\degr$}

\begin{figure}
\begin{center}
\includegraphics[width=0.46\textwidth]{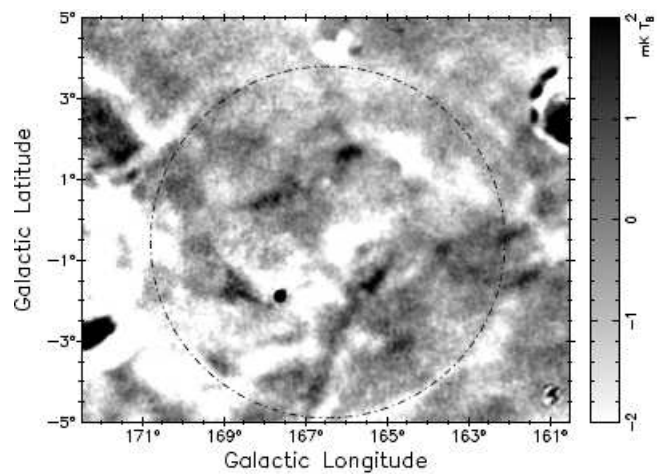}
\caption{$U$ map at $\lambda$6\ cm of the large magnetic bubble. The dotted line outlines the outer boundary of the bubble as quoted by \citet{Kothes04}.}
\label{bubble}
\end{center}
\end{figure}

Two coinciding magnetic bubbles of different sizes were identified in the area $\ell$ around $165\degr$ by \citet{Kothes04} from polarization observations with about 1$\arcmin$ resolution at $\lambda$21\ cm with the DRAO synthesis telescope. These bubbles 
were also seen in the Effelsberg 1.4~GHz polarization data \citep{Reich04} at 9$\farcm4$ resolution.
The smaller bubble extends from $\ell$ = 164$\degr$ to $167\fdg5$ and $-2\degr$ to 2$\degr$ in $b$, while the larger one has about the same projected centre but is approximately 9$\degr$ in size. At $\lambda$6\ cm the two bubbles become very faint. The $\lambda$21\ cm map \citep{Kothes04} was 
limited to $b = -3\degr$, while the $\lambda$6\ cm map shows the larger bubble to extend approximately to $b = -5\degr$ (Figs.~\ref{section2} and~\ref{bubble}). 

The bubbles are faint but visible in the $\lambda$6\ cm Stokes $U$ map  
(Fig.~\ref{bubble}). They are marginally traced in the Stokes $Q$ map, regardless of including large-scale polarized emission or not.
At $\lambda$6\ cm the bubbles can not be identified as discrete objects in $PI$ exceeding just a few times 
the rms-noise value of about 0.4~mK\ $\rm T_{B}$, although
some individual patches or filaments with stronger emission in this large area may be attributed to them. The faintness of the bubbles
 indicates that either their $RM$ is not as high as that of the more pronounced $\lambda$6\ cm Faraday Screens described in this paper, or that 
the polarized background emission
is very faint in respect to the foreground, e.g. in terms of our Faraday Screen model $c$ is close to 1.  

In their preliminary model \citet{Kothes04} discussed the large bubble as a Faraday Screen  with a rather low electron density in agreement with the low H$\alpha$ 
emission in its direction.  Rather important is the association of an \ion{H}{I} bubble identified in the CGPS data with a velocity of about $-20$~km s$^{-1}$  
surrounding the outer larger magnetic bubble, which makes it a Perseus arm object at 1.5~kpc to  2~kpc distance.
This results in a very large object of about 240~pc to 310~pc in diameter. Our $\lambda$6\ cm data will not improve the physical parameters of the bubbles.

\subsubsection{The $\ell = 173\degr$ \ion{H}{II} complex}

\begin{figure}
\begin{center}
\includegraphics[width=0.37\textwidth]{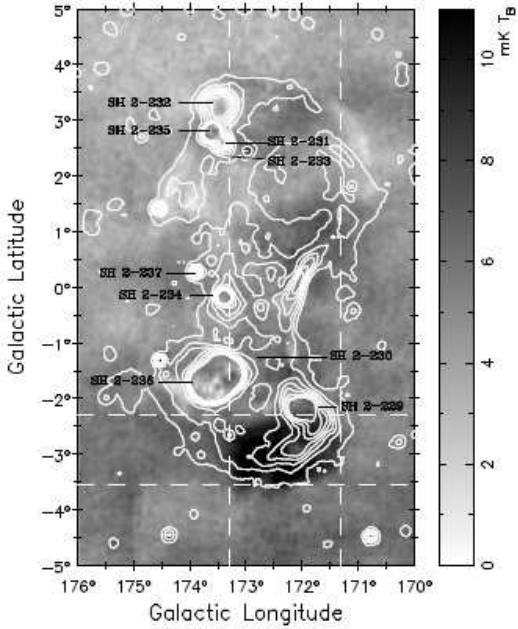}
\caption{$\ell$=173$\degr$ \ion{H}{II} complex. Restored $PI$ in grey scale overlaid by $I$ contours
are shown. The contour lines run from 7~mK $\rm T_{B}$ in steps of 15~mK $\rm T_{B}$ up to 112~mK $\rm T_{B}$.
The two large rectangular regions (vertical and horizontal) marked by dashed lines are the regions where $I$, $PA$ and $PI$ were
averaged as shown in Fig.~\ref{g173cut}. }
\label{g173}
\end{center}
\end{figure}

The survey region around $\ell=173\degr$ is very rich in structures (Fig.~\ref{g173}). Nine \ion{H}{II} regions from the 
Sharpless catalogue are located within this very extended ``bow-tie'' shaped complex 
with the most prominent and extended \ion{H}{II} region, SH 2-236, located in the southeast. 
There are four \ion{H}{II} regions, SH 2-231, SH 2-232, SH 2-233 and SH 2-235 located in the north, among which, 
SH 2-235 is the strongest. The so-called  ``Spider Nebula'' and the ``Fly Nebula'', SH 2-234 and SH 2-237,
 are seen in the middle of the complex, while SH 2-229 is located in the south-west. The central diffuse emission is attributed 
to SH 2-230. All these \ion{H}{II} regions are infrared-bright sources. Besides the Sharpless \ion{H}{II} regions 
additional ridge-shaped structures are seen in the $\lambda$6\ cm $I$ map, for example at
$\ell = 171\fdg20, b = 2\fdg35$ and $\ell = 172\fdg65, b = 3\fdg60$. Thermal characteristics of these filamentary structures are confirmed 
by applying the TT-plot method to derive spectral information between the $\lambda$6\ cm and the corresponding Effelsberg $\lambda$21\ cm 
survey map.

\begin{figure}
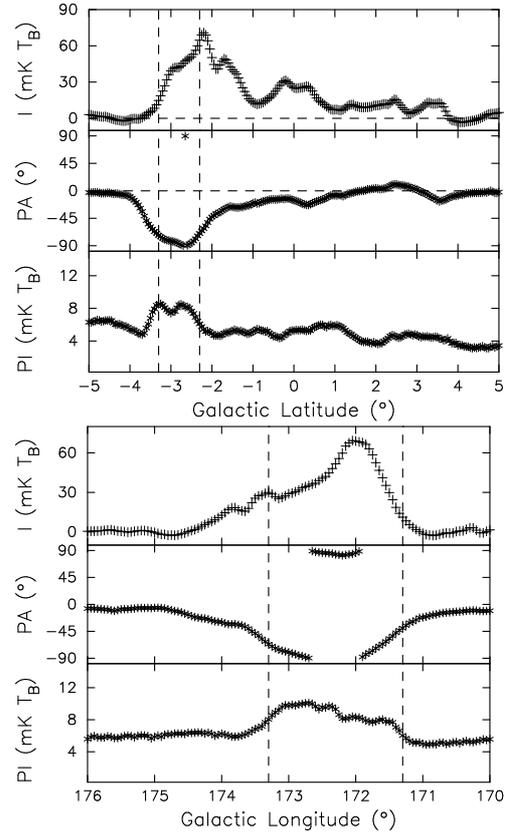

\begin{center}
\includegraphics[angle=-90, width=0.35\textwidth]{13793fig18a.ps}
\includegraphics[angle=-90, width=0.35\textwidth]{13793fig18b.ps}
\caption{Upper plot shows $I$, $PA$ and $PI$ averaged in the vertical rectangle region marked in Fig.~\ref{g173}, where data are averaged in 
columns. The vertical dashed lines mark the boundary of the highly polarized emission region. The lower plot shows the results when averaging in the horizontal rectangle in Fig.~\ref{g173}.}
\label{g173cut}
\end{center}
\end{figure}

\begin{figure}
\begin{center}
\includegraphics[width=0.37\textwidth]{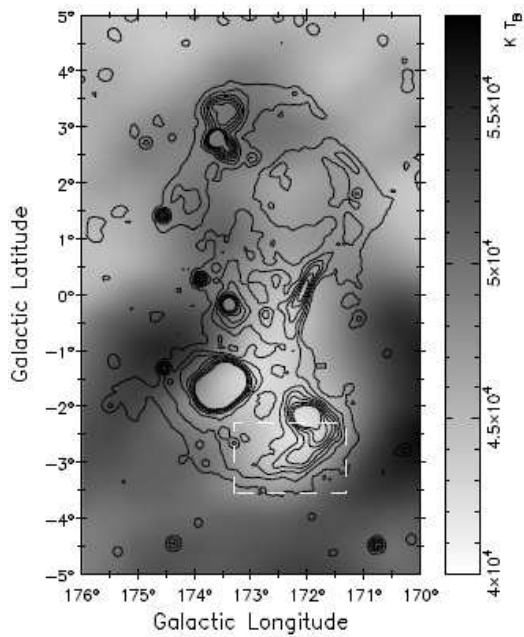}
\caption{22~MHz $I$ map in grey scale of the $\ell$=173$\degr$ \ion{H}{II} complex overlaid by 
$\lambda$6\ cm $I$ contours. The contour intervals are the same as 
in Fig.~\ref{g173}. Strong thermal absorption can be seen in the region with excessive $PI$ at $\lambda$6\ cm.}
\label{g173low}
\end{center}
\end{figure}

\begin{figure}
\begin{center}
\includegraphics[angle=-90, width=0.36\textwidth]{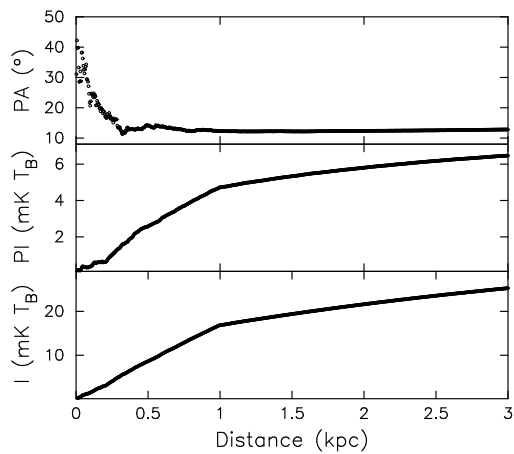}
\caption{Simulated and accumulated $\lambda$6\ cm $PA$, $PI$ and $I$ data as a function of distance in the direction $\ell = 172\fdg3, b = -2\fdg9$. 
The data are averaged within $0\fdg2$.}
\label{g173sim}
\end{center}
\end{figure}

Excessive polarized emission extending for about $2\fdg0 \times 1\fdg3$ is seen in the southern part of the complex, distinct 
from SH 2-229 and SH 2-236 (Fig.~\ref{g173}). $PI$ is about 3~mK $\rm T_{B}$ higher compared to its surroundings (Fig.~\ref{g173cut}). 
The distances to the two Sharpless regions are 510~pc for SH 2-229 and 3.2~kpc for SH 2-236 \citep{Blitz82}. No morphological resemblance exists between the  
polarized patch and total intensity emission, while it seems that the western part of the polarized structure is overlapped with a
southern extension from SH 2-229 (see Fig.~\ref{g173}). This patch coincides with thermal absorption structures visible in the 
low-frequency maps at 10~MHz 
by \citet{Caswell76} and at 22~MHz by \citet{Roger99} as shown in Fig.~\ref{g173low}. Thus we expect thermal emission to act as a Faraday Screen. 
However, we can not apply our simple Faraday Screen model, because the polarized ``$\mathrm{on}$'' emission exceeds the ``$\mathrm{off}$'' emission, which
requires that the background and the foreground $PA$s are different. According to the starlight polarization catalogue by \citet{Heiles00}, large $PA$ variations 
are observed in this direction. High-angular resolution Galactic emission simulations as described by \citet{Sun09} were used to simulate the $\lambda$6\ cm
$PA$, $PI$ and $I$ as a function of distance (Fig.~\ref{g173sim}). A significant change of $PA$ is seen for distances below 300~pc, which is needed to explain 
a $PI$ excess by Faraday rotation. This indicates that the Faraday Screen is nearer than 300~pc. 

Low-frequency absorption is more pronounced by local features with strong emission background than by more distant
objects having the same physical properties. The clear absorption at 10~MHz and 22~MHz of G172.3-2.9 (see Fig.~\ref{g173low}) 
further supports a local origin. The recent 1.4~GHz polarization survey by Landecker et al. (2010, submitted) with 1$\arcmin$
angular resolution does not show a corresponding $PI$ structure, however, the $PA$ distribution is rather smooth in this area.  
Observations at other wavelengths are needed to constrain the properties of this outstanding Faraday Screen in our $\lambda$6\ cm map. 

\section{Summary}

In Paper~II, we present the second section covering the outer Galaxy for the area $129\degr \leq \ell \leq 230\degr$, $|b| \leq 5\degr$ of the Sino-German $\lambda$6\ cm polarization survey 
of the Galactic plane at an angular resolution of $9\farcm5$. 
It is the ground-based polarization survey at the highest frequency for the Galactic 
anti-centre region. The observed polarization data have been restored to an absolute level by adding extrapolated large-scale components from the 
WMAP K-band polarization maps \citep{Hinshaw09}. 

Numerous newly detected Faraday Screens indicate the presence of large magnetic bubbles in the ISM hosting regular magnetic fields of a few $\mu$G.  
A simple model fit to selected Faraday Screens, which also includes \ion{H}{II} regions, was used to estimate their physical parameters. Our main results are:

\begin{enumerate}
 
\item We note that the remarkable polarized ``lens'' Faraday Screen in front of W5 detected by \citet{Gray98} at $\lambda$21\ cm becomes invisible at $\lambda$6 \ cm, 
while previously unknown polarized structures were detected at $\lambda$6\ cm 
at the boundaries of W5.

\item The Faraday Screen model fits for LBN~676, LBN~677 and LBN~679 indicate that besides the established 
Perseus arm objects LBN~676 and LBN~679 also LBN~677 is located in the Perseus arm rather than at 0.8~kpc distance,
because its polarized foreground level is as that of the other two objects. The parameters of the three LBNe are listed in Table~3. 
For LBN~676 the model fit indicates a magnetic field direction opposite to those of the other two LBNe. 
A similar case was noted by \citet{Mitra03} for a few Perseus arm \ion{H}{II} regions based on pulsar $RM$s shining through them. 

\item The newly discovered extended Faraday Screen G146.4-3.0 is likely quite local. 
$B_{\parallel}$ is estimated to be about $-8.6~\mu$G if located at 690~pc, which is most likely an upper limit. The field strength within G146.4-3.0 will increase in case its distance is smaller. 

\item The two huge polarized bubbles located at $\ell = 165\degr$ as revealed by \citet{Kothes04} 
at $\lambda$21\ cm become very faint at $\lambda$6\ cm.

\item An extended blob showing excessive polarized emission is detected in the lower area of the ``bow-tie'' shaped \ion{H}{II} region complex around $\ell = 173\degr$. Absorption at lower radio frequencies coincides with the $PI$ 
excess. We find evidence that this is a local Faraday Screen with a likely distance smaller than 300~pc. 

\end{enumerate}

For most of the selected Faraday Screens the polarized emission becomes weaker compared to their surroundings. 
This is expected when the $PA$ of the background polarization is rotated away from the foreground
direction. The polarized emission exceeds that of the surroundings only in case the 
difference of the $PA$s is reduced.

The selected Faraday Screens from the $\lambda$6\ cm polarization survey demonstrate the existence of numerous high-$RM$ features in the interstellar medium. These structures cover a significant fraction of
the surveyed area. $RM$ studies based on pulsars or extragalactic sources aiming to derive the parameters of the
large-scale Galactic magnetic field need to take the $RM$-contribution from Faraday Screens into account. The formation
of strong regular magnetic fields in thermal low-density regions exceeding the interstellar value needs to be 
investigated. We note that most of the
Faraday Screens visible at $\lambda$6\ cm are not seen at longer wavelengths, where their $RM$ causes polarization
angle rotations exceeding $180\degr$. Missing depolarization implies that small-scale fluctuations across the beam
may not be significant. 

\begin{acknowledgements}
We thank the staff of the Urumqi Observatory of NAOC on the Nanshan mountain for the assistance during the observations and its director,
Prof. Nina Wang for granting observing time. Particularly, we like to thank Mr. Otmar Lochner for constructing and installing the $\lambda$6\ cm 
receiving system at the Urumqi telescope and Dr. Peter M\"uller for the installation and adaptation of Effelsberg data reduction software at the Urumqi observatory. 
The Chinese survey team is supported by the National Natural Science foundation of China (10773016,10833003,10821061) and the National Key Basic Research 
Science Foundation of China (2007CB815403). X. Y. Gao thanks the joint doctoral training plan between CAS and MPG and financial support from CAS and MPIfR for making his stay possible at MPIfR Bonn.
\end{acknowledgements}

\bibliographystyle{aa}
\bibliography{bbfile}

\appendix

\end{document}